\documentclass[superscriptaddress,groupedaddress,nofootnoteinbib,12pt]{article}  % for review and submission
 \pdfoutput=1
\usepackage{graphicx}  % needed for figures
\usepackage{dcolumn}   % needed for some tables
\usepackage{amsmath}                                      
\usepackage{bm}        % for math

\usepackage{color}	
\usepackage{jcappub}
\usepackage[normalem]{ulem}
\usepackage{hyperref}

\def\be{\begin{equation}}
\def\ee{\end{equation}}
\def\ba{\begin{eqnarray}}
\def\ea{\end{eqnarray}}
\def\beq{\begin{eqnarray}}
\def\eeq{\end{eqnarray}}

\def\noi{{\noindent}}
\definecolor{verde}{rgb}{0,0.5,0}
\def\blu{\color{blue}}

\def\L*{{\cal L}_*}
\def\L{\mathcal{L}}
\def\({\left(}
\def\){\right)}

\def\<{\langle}
\def\>{\rangle}

\def\cs2{c_{s}^{2}}

 \def\be   {\begin{equation}}   \def\ee   {\end{equation}}
 \def\ba   {\begin{array}}      \def\ea   {\end{array}}
 \def\bea  {\begin{eqnarray}}   \def\eea  {\end{eqnarray}}
 \def\bean {\begin{eqnarray*}}  \def\eean {\end{eqnarray*}}

\usepackage{xcolor}
\definecolor{MatteoColour}{rgb}{0.,0.7,0.3}
\begin{document}

\title{Non-Gaussianity from Axion-Gauge Fields Interactions during Inflation}

\author{Emanuela Dimastrogiovanni$\,^{a,b}$, Matteo Fasiello$\,^{c}$, Robert J. Hardwick$\,^{c}$, Hooshyar Assadullahi$\,^{c}$, Kazuya Koyama$\,^{c}$, David Wands$\,^{c}$}
%-1.9cm
\affiliation{\hspace{-0.5cm}$^{a}$ CERCA \& Department of Physics, Case Western Reserve University,
Cleveland, OH, 44106, USA.}
\affiliation{\hspace{-0.5cm}$^{b}$ Perimeter Institute for Theoretical Physics,
31 Caroline Street North, Waterloo, N2L 2Y5, Canada.}
\affiliation{\hspace{-0.5cm}$^{c}$ Institute of Cosmology and Gravitation, University of Portsmouth, PO1 3FX, UK.}

\date{\today}

%%%%%%%%%%%%%%%%%%%%%%%%%%%%%%%%%%%%%%%%%%%%%%%%%%%%%%%%%%%%%%%%%%%%%
%%%% Abstract

\abstract{We study the scalar-tensor-tensor non-Gaussian signal in an inflationary model comprising also an axion coupled with SU(2) gauge fields. In this set-up, metric fluctuations are sourced by the gauge fields already at the linear level providing an enhanced chiral gravitational waves spectrum. The same mechanism is at work in generating an amplitude for the three-point function that is parametrically larger than in standard single-field inflation.}

%The bispectrum is thus an important observable, complementary to the power spectrum, for testing the model.

\maketitle

%%%%%%%%%%%%%%%%%%%%%%%%%%%%%%%%%%%%%%%%%%%%%%%%%%%%%%%%%%%%%%%%%%%%%

\section{Introduction}

A period of accelerated expansion in the early universe, \textit{inflation}
%, typically lasting at least 55 e-folds, 
has been hypothesized \cite{inflation} in order to solve a number of puzzling initial conditions in the standard hot big-bang cosmology.  Already in its simplest formulation, that of a scalar field minimally coupled to gravity, inflation can resolve such issues and provides a mechanism by which quantum fluctuations at early times are swept up by inflation to become the primordial seeds for structures to form in the universe.
%\\

The spectacular advances in observational cosmology in recent decades have refined the allowed range for viable inflationary models.  A nearly scale-invariant spectrum of primordial adiabatic scalar fluctuations are required for agreement with observations with only small ($f_{\text{nl}}\lesssim\mathcal{O}(10)$) non-Gaussianities \cite{observations}. These constraints notwithstanding, the inflationary paradigm can accommodate a rich particle content. An observational window on inflation is then automatically also a precious portal to high energy physics and a very special one at that; it provides access to beyond-standard-model energy scales that can be as high as $10^{14} {\rm GeV}$, well out of the reach of earth-bound particle colliders.
%\\ 
Extra field content is not just an interesting possibility for inflation, it is also a natural one. To give just one example, in supersymmetric theories, unless supersymmetry is broken at scales much higher than the Hubble scale, $E\gg H$, the inflationary vacuum energy will break supersymmetry so that some of the resulting particles will have masses $m\sim H$. Even if such massive fields are long lost to us today, cosmological observables (e.g., the squeezed configuration of the bispectrum) can carry the imprint of their early dynamics so that one can engage in cosmological archeology and search for such \textit{fossils} \cite{fossils}. Interestingly, information on the spin, mass and coupling of these particles can still be accessible today \cite{Arkani-Hamed:2015bza}.
%\\

%\indent 
Given the plethora of inflationary setups still compatible with observational bounds, one may rely on future cosmological probes  to identify the most compelling scenarios, as well as the requirement of a theoretically robust implementation of inflation. The latter includes navigating the perils of the so-called $\eta$-problem; in the absence of a sufficiently powerful symmetry, the inflaton potential will receive loop corrections of the form $V_0 (\phi/M_P)^n$ making its mass too large ($m\sim H$) to sustain a sufficiently long expansion.
%\\

An approximately shift-symmetric potential can significantly ameliorate the $\eta$-problem as in the well-known case of natural inflation \cite{naturalinflation}. In this context (see \cite{Pajer:2013fsa} for a review on the subject), the axion potential receives non-perturbative contributions  from the gauge sector resulting in a left-over discrete shift symmetry for the field $\phi \rightarrow \phi + 2\pi f$, with the dimensionful quantity $f$ regulating the periodicity. 
Known string-theory constructions \cite{fmpt} suggest the constraint $f<M_P$; this hierarchy  is further motivated by the fact that quantum gravity is expected to break all global symmetries. Given that observationally viable inflation via a single axion requires $f> M_P$, in order to operate in an under-control inflationary regime one may couple the axion to other\footnote{Another intriguing possibility is to have multiple axions \cite{Kim:2004rp}.} sectors so as to effectively lower $f$. 
%\\

%\noi 
One such example is that of an axionic inflaton directly coupled to gauge fields via the least-irrelevant shift-symmetric operator $\phi F\tilde{F}$. There exists a vast literature \cite{vast} on what remains a very active subject, and includes the possibility of $F\tilde{F}$ being standard model gauge fields thus providing a natural reheating scenario (see e.g. \cite{Ferreira:2017lnd}). In light of the axion-gauge coupling, an entire class of axion inflation models share intriguing potential signatures: a chiral gravitational wave signal, and in particular one that can grow at smaller scales (blue spectrum)\footnote{Other classes of inflationary scenarios endowed with non-standard gravitational waves production mechanisms include, e.g., scalar spectator fields with a small sound speed \cite{cs} and modified gravity models \cite{mg}.}.  It is worth pointing out that similar models have recently been employed in the context
leptogenesis via  axial-gravitational anomaly \cite{Caldwell:2017chz}.%\\

%\indent 
An intriguing specific realization of axion inflation is known as chromo-natural inflation (CNI) \cite{recentCNI}: here the coupling is to SU(2) gauge fields\footnote{See \cite{Lozanov:2018kpk} for a very recent analysis pointing out one extra advantage that comes with the use of an $SU(2)$ as opposed to $U(1)$ model in the context of Schwinger pair creation and backreaction.}, allowing for isotropic background solutions (\cite{Maleknejad:2013npa,Bielefeld:2014nza} provide a non-exhaustive list of works on the subject). Remarkably, a scan of the parameter space of CNI reveals regions generating signatures detectable by both CMB probes and interferometers. Further studies  \cite{Dimastrogiovanni:2012ew} showed that the simplest realization of CNI is excluded by Planck data. This has lead to an extension of the model   \cite{Dimastrogiovanni:2016fuu} (see also \cite{Obata:2016tmo}) that retains all its  original intriguing features; the tension with data is resolved by equipping the scalar sector with an additional field, now driving inflation. The extra field is not necessarily an axion and therefore its potential need not be shift-symmetric. Crucially, detection-level gravitational waves can be generated already at sub-Planckian values for the axion field-excursion, thereby reducing the effect of loop corrections on the inflationary potential. Interestingly, it has recently been shown how both CNI and its extension can be embedded in supegravity and string theory \cite{DallAgata:2018ybl}.
%\\
%\indent 

We consider in this paper the model introduced in \cite{Dimastrogiovanni:2016fuu}. The SU(2)-based enhancement of  gravitational waves  can lead to detectable $\langle EB \rangle$, $\langle TB \rangle$ signals for upcoming CMB probes and may be searched for using existing interferometers (and a cross-correlation thereof) \cite{Thorne:2017jft}. It has recently been shown that this setup supports large tensor non-Gaussianities \cite{Agrawal:2017awz} and we will show here that the same is true for the scalar-tensor-tensor correlation.\\\indent  This paper is organized as follows: in \textit{Section} \ref{model} we review the model and its predictions at the level of the power spectra; in \textit{Section} \ref{bispectrum} we present the calculation of the scalar-tensor-tensor bispectrum and discuss our findings on its shape and amplitude with an eye on perturbativity bounds inherited also from the scalar-sector;  conclusions are  in \textit{Section} \ref{conclusions}. More details about the calculations can be found in the Appendices.

\section{The model}
\label{model}

\indent As mentioned above, our model includes \textsl{spectator} fields (i.e.~fields providing a sub-leading contribution to the total energy density during inflation), including an axion  field $\chi$ and an SU(2) gauge field $A_{\mu}^{a}$, in addition to the inflaton sector,
\begin{equation}
\mathcal{S}=\int d^4x\sqrt{-g}\left[\frac{M_{\text{Pl}}^{2}}{2}R+\mathcal{L}_{\phi}-\frac{1}{2}\left(\partial\chi\right)^{2}-U(\chi)-\frac{1}{4}F_{\mu\nu}^{a}F^{a\mu\nu}+\frac{\lambda\,\chi}{4f}F_{\mu\nu}^{a}\tilde{F}^{a\mu\nu}\right]\,,
\end{equation}
where $\mathcal{L}_{\phi}$ is the inflaton Lagrangian, $F_{\mu\nu}^{a}\equiv\partial_{\mu}A^{a}_{\nu}-\partial_{\nu}A^{a}_{\mu}-g ^{abc} A^{b}_{\mu}A_{\nu}^{c}$ and the definition defined $\tilde{F}^{a\mu\nu}\equiv\epsilon^{\mu\nu\rho\sigma}F^{a}_{\rho\sigma}/(2\sqrt{-g})$ has been used. 
%\\ 

The background equations of motion and the linear perturbation analysis were first presented in \cite{Dimastrogiovanni:2016fuu}. In this section, we review the main results and identify the model parameters that will appear in the bispectrum computation. The background for the gauge field can be chosen as $A_{0}^{a}=0$, $A_{i}^{a}=\delta_{i}^{a} a(t) Q(t)$. The scalars $Q$ and $\chi$ have coupled equations of motion. Under minimal assumptions on the parameters and in a regime of slow-roll for the fields, the effective potential for $Q$ is minimized by 
\begin{equation}\label{sol-sr1}
Q_{}=\left(\frac{-f\, U_{\chi}}{3g\lambda H}\right)^{1/3}\,.
\end{equation}
From the same equations of motion it also follows that
\begin{equation}\label{sol-sr2}
\frac{\lambda}{2fH}\dot{\chi}\simeq m_{Q}+\frac{1}{m_{Q}}\,,
\end{equation}
where the parameter $m_{Q}\equiv g\,Q/H$ is to be interpreted as the mass, in units of Hubble, of the gauge field fluctuations. Einstein's equations lead to the following relation among slow-roll parameters:
\bea \epsilon_H \equiv -\dot{H}/H^2=\epsilon_{\phi}+\epsilon_{\chi}+\epsilon_{B}+\epsilon_{E}\; ,
\eea
\noi with 
\bea
\epsilon_{\phi}\equiv\frac{\dot{\phi}^{2}}{(2H^2 M_{\text{Pl}}^{2})}\,;\;\;\;
 \epsilon_{\chi}\equiv\frac{\dot{\chi}^{2}}{(2 H^2 M_{\text{Pl}}^{2})}\,;\;\;\; \epsilon_{B}\equiv \frac{g^2 Q^4}{(H M_{\text{Pl}})^{2}}\,;\;\;\;  \epsilon_{E}\equiv \frac{(HQ+\dot{Q})^2}{(HM_{\text{Pl}})^2} \; .
\eea

%\noindent 
The metric tensor fluctuations ($h_{ij}$)  are linearly sourced by the tensor perturbations of the gauge field. The latter experience (near horizon crossing) a growth in one of their two polarizations that is controlled by $m_{Q}\,$; as a result, the corresponding helicity in the gravitational waves is enhanced. This non-zero chirality can be understood as a consequence of the parity-breaking nature of the gauge-field background. The expression for the sourced power spectrum is given by
\begin{equation}\label{psh}
\mathcal{P}^{\text{s}}_{h}= \epsilon_{B}\,\frac{H^{2}}{\pi^2 M_{Pl}^{2}}\mathcal{F}^{2}\,,
\end{equation}
where $\mathcal{F}= \mathcal{F}(m_{Q}) $ (a more detailed derivation can be found in \cite{Dimastrogiovanni:2016fuu}). The transient instability of the gauge field tensor fluctuations can be understood as an energy transfer from the rolling axion.  
%\\

The $SU(2)$-sensitive contribution to the power spectrum of  gravitational waves (GW) can be larger than the one from vacuum fluctuations and  within reach of upcoming experimental probes \cite{Dimastrogiovanni:2016fuu}; the model predicts chiral gravitational waves that would be observable
for a sizable portion of its parameter space \cite{Thorne:2017jft}. Our set-up serves as an explicit example of the fact that detectable GW may be generated even at a relatively low value for $H$, thus breaking the one-to-one $r \leftrightarrow H$ correspondence between the tensor-to-scalar ratio and the energy scale of inflation (see \cite{Fujita:2017jwq} for more about the lower bound on $H$ in this context).
%\\

The tensor power spectrum is characterized by a  broad (depending on model parameters) feature, a distinctive scale dependent ``bump" that  results from  the background evolution of the axion-gauge field system. From this feature originates the fact that there is in this model ample room for a blue tensor spectral index, crucial for direct detection by interferometers. Having lifted the burden of driving inflation from the axion (in order to recover compatibility with data \cite{Dimastrogiovanni:2012ew}), in the extended model one may enhance sourced gravitational waves on different scales by sampling the parameter space and acting on the coupling  between the fields. 
%\\   

%\noindent 
Moving on to the power spectrum of curvature fluctuations, this will depend on the precise form of $\mathcal{L}_{\phi}$ and, naturally, also on the specific dynamics one may postulate for the post-inflationary evolution of the spectator sector. The axion $\delta\chi$ and gauge field $(\delta Q,\,M)$ scalar fluctuations are directly coupled with one another (and only gravitationally coupled to the inflaton fluctuations). We note that these modes will undergo a tachyonic instability \cite{Dimastrogiovanni:2012ew} starting in the sub-horizon regime unless $m_{Q}\geq\sqrt{2}$. We will confine our analysis to such viable region of the parameter space.
%\\

Both the inflaton field and scalar fluctuations of the spectator sector contribute to curvature perturbations. The authors of \cite{Dimastrogiovanni:2016fuu} chose to be as agnostic as possible on the details of $\mathcal{L}_{\phi}$. It is nevertheless necessary to ensure that the field $\phi$ is the one driving inflation and that furthermore there exists a hierarchy among the slow-roll parameters with $\epsilon_{\phi}\simeq\epsilon_{\rm max}$. The latter condition ensures that the spectral index can satisfy existing observational constraints.  Under such conditions the power spectrum of curvature perturbation is dominated by the inflaton contributions and is only mildly affected by the axion and gauge fields. However, for a more careful analysis, see \textit{Section} \ref{results}.  
% \\

%but at the level of the overall amplitude and of the spectral index. If one relaxes these, minimal, assumptions a richer phenomenology arises that may well affect the predictions for the scalar power spectrum, possibly introducing effects on small scales.

%\noi {\verde - cite also most recent by Eiichiro on Schwinger either here or in the introduction}

\section{STT bispectrum %$\langle \zeta\, h\,h \rangle$
 from Chern-Simons interactions}
\label{bispectrum}

\noindent The $\chi\, F\tilde{F}$ interaction of Eq.~(\ref{model}) supports a transient  growth in one of the $SU(2)$ tensor polarizations that propagates to the corresponding helicity in the GW power spectrum. This mechanism is also in place for higher-order correlation functions. The GW bispectrum for the theory in Eq.~(\ref{model})  has been calculated in \cite{Agrawal:2017awz}, where it was shown that the $SU(2)$ contribution to tensor non-Gaussianity can be significantly larger than the one of standard single field inflation.

%\noi 
It is intuitively clear that, because the growth of the sourcing mode function occurs (only) near horizon crossing, the bispectrum shape will very much resemble the equilateral one, although there are some subtle differences with respect to the exact equilateral template. In an analogous fashion, one expects also mixed tensor-scalar correlators to receive the most sizable contributions from the gauge sector in equilateral configurations. The axion $\delta\chi$ is sourced by gauge tensor fluctuations via Chern-Simons interactions while in turn the curvature perturbation $\zeta$ receives contributions from $\delta\chi$.
%\\

%\noindent 
Scalar non-Gaussianity is constrained on large CMB scales by $f_{\text{nl}}\lesssim\mathcal{O}(10)$ \cite{observations}. These bounds will soon improve thanks to upcoming large-scale structure observations and new CMB polarization data. The ongoing development of new interferometers with improved sensitivity to the stochastic background of primordial GW \cite{lisa} will also help us place stronger constraints on tensor and mixed non-Gaussianity. Non-Gaussian observables are invaluable as a probe of the production mechanism of primordial GW and, more broadly, the inflationary particle content \cite{fossils}.
%\\ 

\begin{figure}[h!]
\begin{center}
  \includegraphics[width=5.5cm]{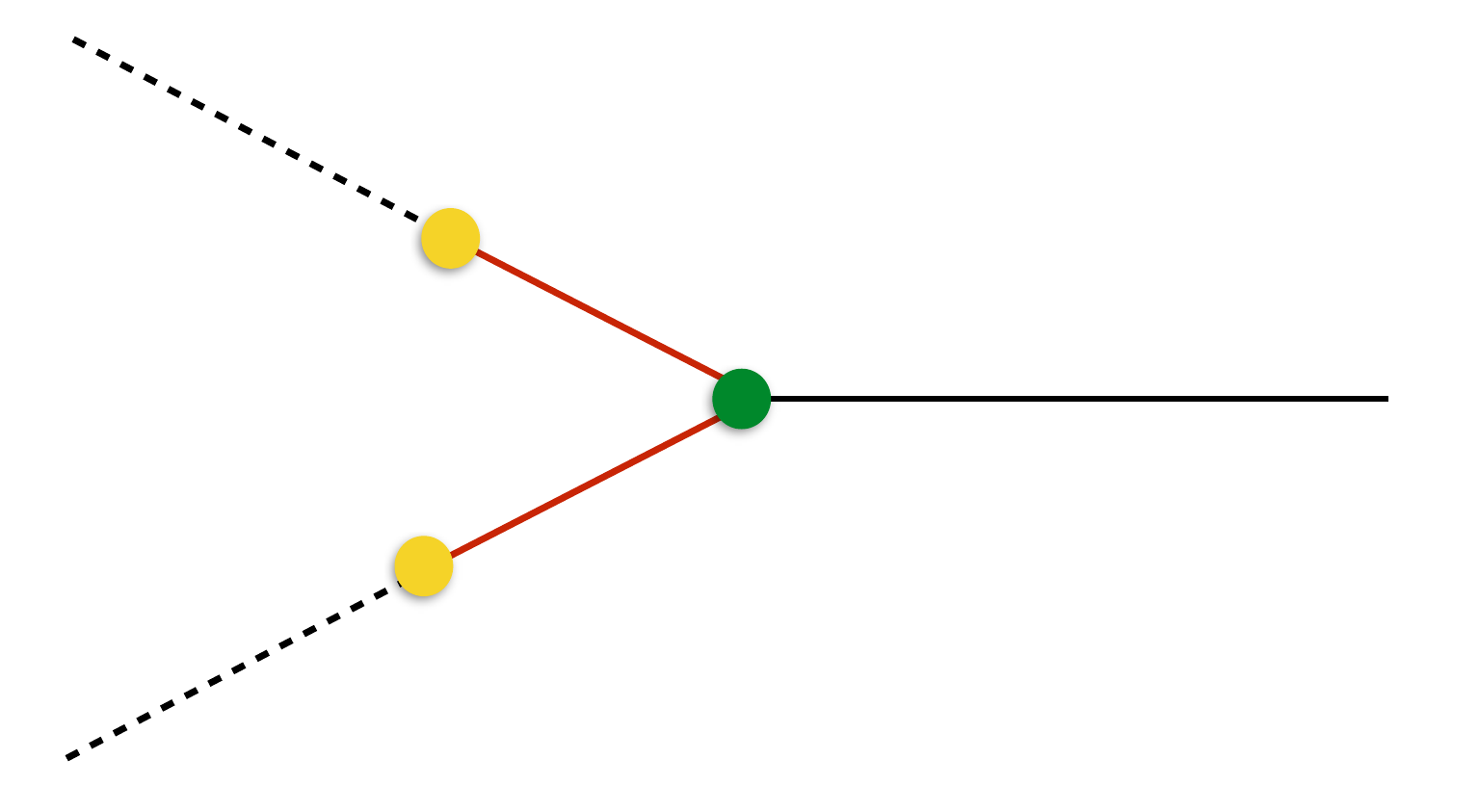}
\caption{Black dotted lines represent the metric tensor $h\,$; red lines stand for the gauge field tensor perturbation $t\,$; the solid black line indicates the curvature perturbation $\zeta^{(\chi)}\,$. The green vertex arises from the Chern-Simons contribution to $\delta\chi\, t \, t$. The yellow vertex (to be understood according to the caveat stressed above) arises from the quadratic mixing term $h\,t$ in the Lagrangian.
}
 \label{fig1}
\end{center}
\end{figure}

%\noindent 
In what follows, we shall focus on the the $(\chi,t)$-mediated $\langle h\,h\,\zeta \rangle$ bispectrum contribution. The presence and form of the Chern-Simons interaction suggest that this one observable is particularly sensitive to the effects of gauge fields. A typical contribution of this kind is represented\footnote{It is important to note here that the diagram in Fig.~\ref{fig1} is meant as a pictorial reminder of the fields and interactions in play but should not be intended as the exact \textit{in-in} formalism representation of the calculation. This is because the presence of the two-fields vertex in yellow requires a specific hierarchy between the interaction it represents and the rest of the quadratic action: $\delta\mathcal{L}^{(2)}_{\rm yellow}\ll \mathcal{L}^{(2)}_{\rm rest}$. Such inequality is not satisfied at all times therefore a  consistent calculation entails either diagonalizing the system to avoid quadratic interactions or the use of Green's functions methods. We adopt the latter. }
in the diagram of Fig.~(\ref{fig1}), which we evaluate in details in the remaining of this section.
%
%
%[ht!]

\noi We will henceforth work with comoving fields, $\Psi_{ij}\equiv a\,M_{Pl}\,h_{ij}/2$, and $u\equiv a\, \delta\chi$. The relation between the comoving curvature perturbation $\zeta$ and the axion field fluctuations, $\delta\chi$  is given, at leading order in $\dot{\chi}/\dot{\phi}$, by
\begin{equation}\label{u-chi-zeta-relation}
\zeta^{\chi}= \frac{U_{\chi}}{V_{\phi}}\left(\frac{H}{\dot{\phi}}\right)\delta\chi\,,
\end{equation}
where $U_{\chi}\equiv\partial_{\chi}U$ and $V_{\phi}\equiv\partial_{\phi}V$. The scalar-tensor-tensor bispectrum then reads 
\begin{equation}
\langle h_{\textbf{p}}(\tau)\,h_{\textbf{q}}(\tau)\,\zeta_{\textbf{k}}^{\chi} (\tau)\rangle =-\frac{4}{a^3(\tau)M_{Pl}^{2}}\left(\frac{U_{\chi}}{V_{\phi}}\right)\left(\frac{H}{\dot{\phi}}\right)\langle \Psi^{}_{\textbf{p}} (\tau)\,\Psi^{}_{\textbf{q}}(\tau)\, u_{\textbf{k}}(\tau)\rangle\,.
\end{equation}

\subsection{Perturbative solutions}
\noi Tensor perturbations are expanded in Fourier space as
\begin{equation}
\mathcal{T}_{ij}(\textbf{x},t)=\int\frac{d^3 k}{(2\pi)^3}\,e^{i\textbf{k}\cdot\textbf{x}}\sum_{\lambda=R,L}e^{\lambda}_{ij}(\hat{k})\,\mathcal{T}_{\textbf{k}}^{\lambda}(t)\,,
\end{equation}
where $\mathcal{T}_{\textbf{k}}^{\lambda}(t)=\mathcal{T}_{\textbf{k}}^{(1)\lambda}(t)+\mathcal{T}_{\textbf{k}}^{(2)\lambda}(t)+\dots$, and dots indicate higher-order terms in the perturbative expansion. In the equation just above $\mathcal{T}$ is a placeholder for $\Psi_{ij}$ as well as for the transverse and traceless part of the gauge field fluctuations $\delta A_{i}^{a}\supset t_{ai}$. This expansion will be convenient in light of the Green's function method, which we adopt throughout this manuscript\footnote{Alternatively, one may switch to a different basis to decouple, up to second order, $SU(2)$ fields from standard tensor modes and then employ the in-in formalism. The two approaches are equivalent.}. Similarly, for the scalar field we have
\begin{equation}
u(\textbf{x},t)=\int\frac{d^3 k}{(2\pi)^3}\,e^{i\textbf{k}\cdot\textbf{x}}u_{\textbf{k}}^{}(t)\,,
\end{equation}
with $u_{\textbf{k}}^{}(t)=u_{\textbf{k}}^{(1)}(t)+u_{\textbf{k}}^{(2)}(t)+\dots$ . To lowest order in the perturbative expansion, one finds
\begin{equation}\label{eq2}
\langle \Psi^{}_{\textbf{p}} \,\Psi^{}_{\textbf{q}}\, u_{\textbf{k}}\rangle = \langle \Psi^{(1)}_{\textbf{p}} \,\Psi^{(1)}_{\textbf{q}} \,u^{(2)}_{\textbf{k}}\rangle+\langle \Psi^{(1)}_{\textbf{p}} \,\Psi^{(2)}_{\textbf{q}} \,u^{(1)}_{\textbf{k}}\rangle+\langle \Psi^{(2)}_{\textbf{p}} \,\Psi^{(1)}_{\textbf{q}} \,u^{(1)}_{\textbf{k}}\rangle\,.
\end{equation}
We are interested in the non-Gaussianity arising from the gauge field effect on the metric tensor perturbations. The latter are linearly sourced by the $SU(2)$ tensor fluctuations, with one helicity acquiring a larger amplitude than the other, as reviewed in \textit{Section}~\ref{model}. We will focus here on the leading helicity mode, setting $\Psi=\Psi^{R}$ from now on.\\
\noi  Formally, the equation of motion for the metric tensor fluctuations reads %(to all orders in perturbation theory)
\begin{equation}\label{eq3a}
\mathcal{O}_{\Psi}\Psi_{\textbf{k}}=\mathcal{S}^{\Psi}_{\textbf{k}}\,,
\end{equation}
where $\mathcal{O}_{\Psi}$ is the operator describing the homogeneous equation of motion, $\mathcal{O}_{\Psi}\Psi_{\textbf{k}}=0$, while $\mathcal{S}^{\Psi}$ acts as a source due to self-couplings as well as to interactions with other fields. The solution to (\ref{eq3a}) {at} $i^{\text{th}}$ order takes the form
\begin{equation}\label{}
\Psi_{\textbf{k}}^{(i)}(\tau)=\int_{\infty}^{\tau} d\tau' \,\mathcal{G}_{\textbf{k}}^{\Psi}(\tau,\tau')  \,\mathcal{S}^{(i)\Psi}_{\textbf{k}}(\tau')\,,
\end{equation}
where $\mathcal{G}_{\textbf{k}}^{\Psi}$ is the Green's function of $\Psi$ and $\mathcal{S}^{(i)\Psi}_{\textbf{k}}$ the source term, expanded {at} the same order. The leading-order terms relevant for the diagram in Fig.~\ref{fig1}, and included in the expansion (\ref{eq2}), are
\begin{eqnarray}\label{eq4}
&&\Psi_{\textbf{k}}^{(1)}(\tau)=\int d\tau' \,\mathcal{G}_{\textbf{k}}^{\Psi}(\tau,\tau')  \,\mathcal{S}^{[\Psi^{(1)} \cdot t^{(1)}]}_{\textbf{k}}(\tau')\,,\\
&&\Psi_{\textbf{k}}^{(2)}(\tau)=\int d\tau' \,\mathcal{G}_{\textbf{k}}^{\Psi}(\tau,\tau')  \,\mathcal{S}^{[t^{(2)}]}_{\textbf{k}}(\tau')\,.
\end{eqnarray}
The quantity $\mathcal{S}^{[\Psi^{(1)} \cdot t^{(1)}]}_{\textbf{k}}$ originates from the quadratic interactions between tensor modes in the metric and the fluctuations of the gauge fields. $\mathcal{S}^{[t^{(2)}]}_{\textbf{k}}$ is the source term for $\Psi$ due the second order perturbation in the gauge field, specifically the one corresponding to the terms $\delta\chi^{}\, t^{2}$ in the cubic Lagrangian. More explicitly, for the gauge field one has
\begin{equation}\label{}
\mathcal{O}_{t} \,t_{\textbf{k}}=\mathcal{S}^{t}_{\textbf{k}}\,,
\end{equation}
and 
\begin{equation}\label{}
t_{\textbf{k}}^{(i)}(\tau)=\int d\tau' \,\mathcal{G}_{\textbf{k}}^{t}(\tau,\tau')  \,\mathcal{S}^{(i)t}_{\textbf{k}}(\tau')\,,
\end{equation}
where $\mathcal{G}_{\textbf{k}}^{t}$ is the Green's function for $t$. The relevant contribution to $t^{(2)}$ is given by 
\begin{equation}\label{}
t_{\textbf{k}}^{(2)}(\tau)=\int d\tau' \,\mathcal{G}_{\textbf{k}}^{t}(\tau,\tau')  \,\mathcal{S}^{[t^{(1)}\cdot u^{(1)}]}_{\textbf{k}}(\tau')\,.
\end{equation}
Once free fields are quantized, the corresponding sourced fields inherit the same set of creation/annihilation operators. {Indicating} $t_{k}^{\lambda}$ as the solution to the homogenous equation of motion for the $SU(2)$ tensor modes\footnote{It can be shown that the homogeneous solution for $t$ is a good approximation for the full 1st order solution up to late times \cite{Dimastrogiovanni:2016fuu}.}, one may write the field operator as
\begin{eqnarray}\label{}
t_{\textbf{k}}^{(1)\lambda}(\tau)=a_{\textbf{k}}^{\lambda}\,t_{k}^{\lambda}(\tau)+a_{-\textbf{k}}^{\lambda\dagger}\,[t_{k}^{\lambda}(\tau)]^{*}\,,
\end{eqnarray}
where $[a^{\lambda_{1}}_{\textbf{k}_{\textbf{1}}},a^{\lambda_{2}\dagger}_{-\textbf{k}_{\textbf{2}}}]=\delta_{\lambda_{1}\lambda_{2}}\,\delta^{(3)}(\textbf{k}_{\textbf{1}}+\textbf{k}_{\textbf{2}})$.\\

\noindent Given the equation of motion for the scalar field, $\mathcal{O}_{u} \,u_{\textbf{k}}=\mathcal{S}^{u}_{\textbf{k}}$, one derives
\begin{equation}\label{}
u_{\textbf{k}}^{(i)}(\tau)=\int d\tau' \,\mathcal{G}_{\textbf{k}}^{u}(\tau,\tau')  \,\mathcal{S}^{(i)u}_{\textbf{k}}(\tau')\,,
\end{equation}
with $\mathcal{G}_{\textbf{k}}^{u}$ the Green's function for $u$. With $u^{(1)}$ the free-field, the expression for $u^{(2)}$ is given by 
%
%\footnote{The axion is sourced, linearly, by the other field fluctuations; these produce additional contributions to tensor-tensor-scalar non-Gaussianity. \textcolor{magenta}{mention them, specifying that our purpose is not to compute the full bispectrum but rather to obtain a }.}
%
%
\begin{equation}\label{}
u_{\textbf{k}}^{(2)}(\tau)=\int d\tau' \,\mathcal{G}_{\textbf{k}}^{u}(\tau,\tau')  \,\mathcal{S}^{[t^{(1)}\cdot t^{(1)}]}_{\textbf{k}}(\tau')\,,
\end{equation}
where $\mathcal{S}^{[t^{(1)}\cdot t^{(1)}]}_{\textbf{k}}$ is, once again, obtained from the cubic Lagrangian in $\delta\chi\,t^{2}$.\\

\noi Let us begin by focusing on $ \langle \Psi^{(1)}_{\textbf{p}} \,\Psi^{(1)}_{\textbf{q}} \,u^{(2)}_{\textbf{k}}\rangle$. To derive $u^{(2)}$, one expands the Chern-Simons interaction to third order %in $(\delta\chi\cdot t \cdot t)$
\begin{equation}\label{eq1}
\mathcal{S}_{\text{CS}}^{(3)}=\frac{\lambda}{2f} \int d^4 x \left\{ \textcolor{black}{-}\,g\,a\,Q\,\delta\dot{\chi}\,\left(t_{ij}\right)^{2}+2\,\delta\chi\,\dot{t}_{ia}\,\epsilon^{0ijk}\,\partial_{j}t_{ka}\right\}\,.
\end{equation}
The corresponding equation of motion for $u$ reads (more details on the derivation can be found in Appendix~\ref{A})
\begin{eqnarray}\label{eq6}
u^{''}_{\textbf{k}}+\left(a^2 m_{\chi}^{2}+k^{2}-\frac{a^{''}}{a}\right) u_{\textbf{k}}&=&\frac{\lambda}{2 f}\int \frac{d^3 k_{2}}{(2\pi)^3}\sum_{\lambda_{1}\lambda_{2}}e_{ij}^{\lambda_{1}}(\hat{k}_{1})e_{ij}^{\lambda_{2}}(\hat{k}_{2})\nonumber\\&\times&\Big\{\textcolor{black}{}aH\Big[ g\,\sqrt{\epsilon_{E}}\,M_{Pl} t^{\lambda_{1}}_{\textbf{k}_{\textbf{1}}}\,t^{\lambda_{2}}_{\textbf{k}_{\textbf{2}}}\nonumber\\&+&m_{Q}\frac{d}{dt}\left(t^{\lambda_{1}}_{\textbf{k}_{\textbf{1}}}\,t^{\lambda_{2}}_{\textbf{k}_{\textbf{2}}}\right)\Big]\textcolor{black}{-}2\,k_{2}\,\frac{d t_{\textbf{k}_{\textbf{1}}}^{\lambda_{1}}}{dt}\,t_{\textbf{k}_{\textbf{2}}}^{\lambda_{2}}\Big\}\,.
\end{eqnarray}
The Green's function for $u$ in the limit of negligible mass for the axion and in the regime $k\tau\rightarrow 0$, is given by
\begin{eqnarray}\label{}
\mathcal{G}^{u}_{\textbf{k}}(\tau,\tau')=\frac{\theta(\tau-\tau')}{k^3\,\tau\,\tau'}\left(k\,\tau' \cos k\tau'-\sin k\tau'\right)\,.
\end{eqnarray}
Notice that, in the massless limit for the axion, one has $\mathcal{G}^{u}_{\textbf{k}}=\mathcal{G}^{\Psi}_{\textbf{k}}$. The final expression for $u^{(2)}_{\textbf{k}}$ is 
\begin{eqnarray}\label{eq44}
u^{(2)}_{\textbf{k}}(\tau)=\int d\tau'\,\mathcal{G}^{u}_{\textbf{k}}(\tau,\tau')\,\mathcal{J}_{\textbf{k}}(\tau')\,,
\end{eqnarray}
where $\mathcal{J}_{\vec{k}}$ is the right-hand side of Eq.~(\ref{eq6}).  Combining Eqs.~(\ref{eq4}) and (\ref{eq44}), after summing over all permutations, one obtains the final result
\begin{eqnarray}\label{20}
 \langle \Psi^{(1)R}_{\textbf{p}} \,\Psi^{(1)R}_{\textbf{q}} \,u^{(2)}_{\textbf{k}}\rangle&\simeq &(2\pi)^3 \delta^{(3)}(\textbf{p}+\textbf{q}+\textbf{k})\left(\frac{\textcolor{black}{-}\lambda}{2\,f}\right)e^{R}_{ij}(-\textbf{p})e^{R}_{ij}(-\textbf{q})\int d\tau^{'}\mathcal{G}_{\textbf{p}}^{\Psi}(\tau,\tau^{'})\,D_{p}(\tau^{'})\nonumber\\&\times&\int d\tau^{''}\mathcal{G}_{\textbf{q}}^{\Psi}(\tau,\tau^{''})\,D_{q}(\tau^{''})\int d\tau^{'''}\mathcal{G}_{\textbf{k}}^{u}(\tau,\tau^{'''})\Big\{\mathcal{A}(p,q,\tau^{'},\tau^{''},\tau^{'''})\nonumber\\&+&\mathcal{B}(p,q,\tau^{'},\tau^{''},\tau^{'''})+\mathcal{C}(p,q,\tau^{'},\tau^{''},\tau^{'''})\Big\}\,,
\end{eqnarray}
where $D_{p}$ is the differential operator defining the quadratic mixing between tensor modes of the metric and tensor perturbations of the gauge field, $D_{p}(\tau)\equiv \frac{2\sqrt{\epsilon_{B}}}{m_{Q}\tau}\partial_{\tau}+ \frac{2\sqrt{\epsilon_{B}}}{\tau^2}(m_{Q}+p\,\tau)$. Note that in writing Eq.~(\ref{20}) we defined
\begin{align}\label{AA}
\mathcal{A}(p,q,\tau^{'},\tau^{''},\tau^{'''})\equiv\sqrt{\epsilon_{E}}\,g\,\,M_{Pl}\,\frac{2}{\tau^{'''}}\text{Re}\Big\{ t_{p}^{}(\tau^{'})t_{q}^{*}(\tau^{''})t_{p}^{*}(\tau^{'''})t_{q}^{}(\tau^{'''})\quad\quad\quad\quad\quad\\\quad\quad\quad\quad\quad\quad\quad\quad+2\,t_{p}^{}(\tau^{'})t_{q}^{}(\tau^{''})t_{p}^{*}(\tau^{'''})t_{q}^{*}(\tau^{'''})          \Big\}  \,,
\nonumber
\end{align}
\begin{align} \label{BB}
\mathcal{B}(p,q,\tau^{'},\tau^{''},\tau^{'''})\equiv -H\,m_{Q}\, \text{Re}\Big\{ t_{p}^{}(\tau^{'})t_{q}^{*}(\tau^{''})\left[t_{p}^{'*}(\tau^{'''})t_{q}^{}(\tau^{'''})+t_{p}^{*}(\tau^{'''})t_{q}^{'}(\tau^{'''})\right]\\\quad\,\,\quad\quad\quad +2\,t_{p}^{}(\tau^{'})t_{q}^{}(\tau^{''})\left[t_{p}^{'*}(\tau^{'''})t_{q}^{*}(\tau^{'''})+t_{p}^{*}(\tau^{'''})t_{q}^{'*}(\tau^{'''})\right]          \Big\}  \,,\nonumber
\end{align}
\begin{eqnarray}\label{CC}
\mathcal{C}(p,q,\tau^{'},\tau^{''},\tau^{'''})&\equiv& -2\,H\,\tau^{'''}  \text{Re}\Big\{ t_{p}^{}(\tau^{'})t_{q}^{*}(\tau^{''})\,q\,t_{p}^{'*}(\tau^{'''})t_{q}^{}(\tau^{'''})\\&&\quad\quad\quad\quad\quad\quad\quad\quad+t_{p}^{}(\tau^{'})t_{q}^{*}(\tau^{''})\,p\,t_{q}^{'}(\tau^{'''})t_{p}^{*}(\tau^{'''})\quad\quad\quad\quad\quad\quad\nonumber\\&&\quad\quad\quad\quad\quad+
2\,t_{p}^{}(\tau^{'})t_{q}^{}(\tau^{''})\,q\,t_{p}^{'*}(\tau^{'''})t_{q}^{*}(\tau^{'''})\nonumber\\&&\quad\quad\quad\quad\quad+2\,t_{p}^{}(\tau^{'})t_{q}^{}(\tau^{''})\,p\,t_{q}^{'*}(\tau^{'''})t_{p}^{*}(\tau^{'''})   \Big\}   
 \,,\nonumber
\end{eqnarray}
where ``${\rm Re}$'' stands for real part and the index R on the mode function for the gauge fields has been dropped for simplicity.\\
 \noindent Before proceeding any further, one ought to point out that the amplitudes stemming from the three contributions in Eq.~(\ref{eq2}) are all parametrically similar to one another (see Eq.~\ref{inspec}). However, the structure of the contributions with $\Psi^{(2)}$ in Eq.~(\ref{eq2}) is different from those with $u^{(2)}$ in that they entail a double time-integral rather than products of independent integrals (see Appendix~\ref{B} for the explicit expressions). Nevertheless, for the reasons outlined above, we expect a similar shape, i.e. with a peak in the equilateral configuration. In the remainder of the section we focus on the $u^{(2)}$ contribution, but we stress that the discussion on the final results applies to both contributions.

\subsection{Amplitudes and shapes}
\label{results}

\begin{figure}[h!]
\begin{center}
  \includegraphics[width=7.2cm]{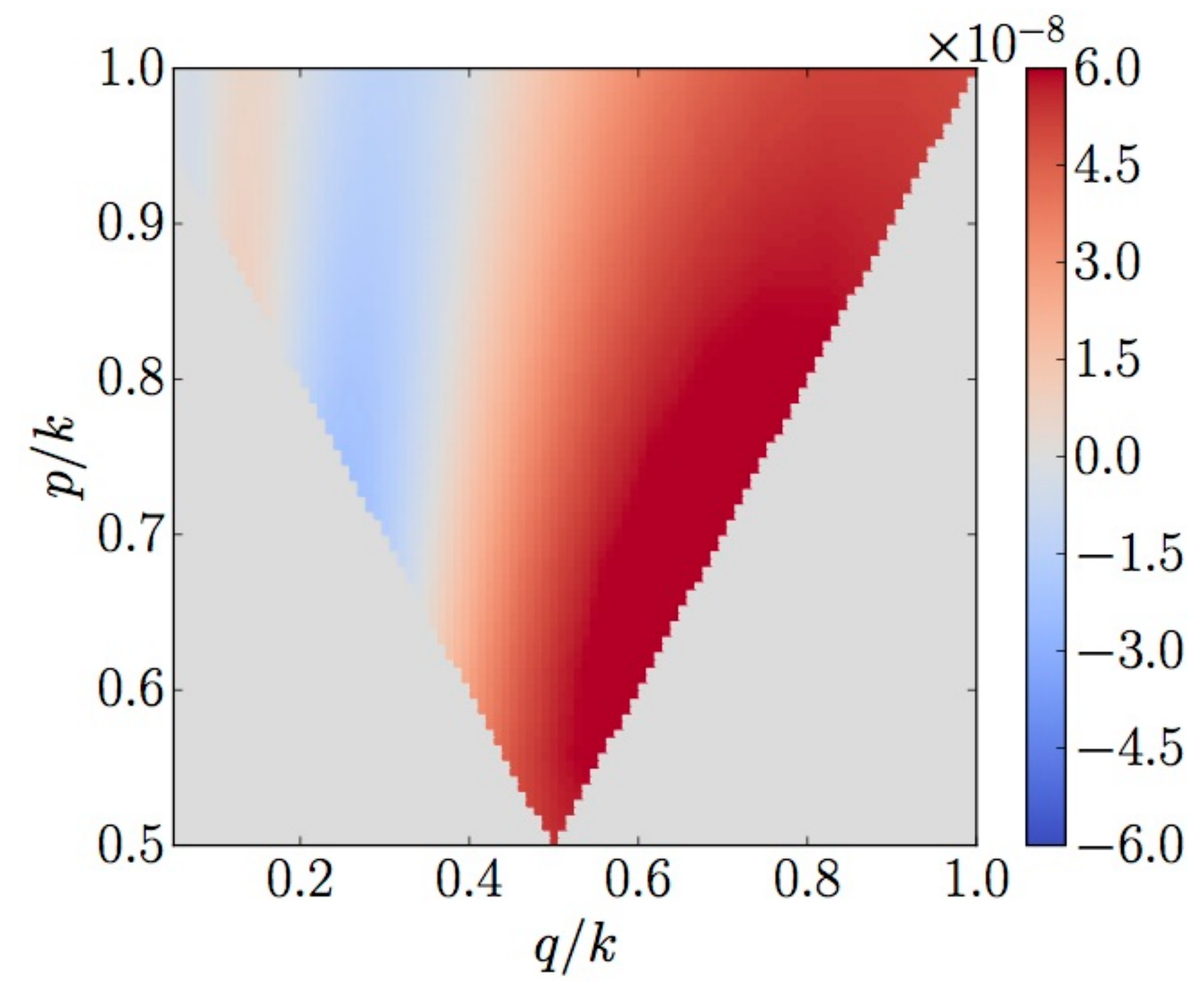} \hspace{0.4cm}
  \includegraphics[width=7.2cm]
{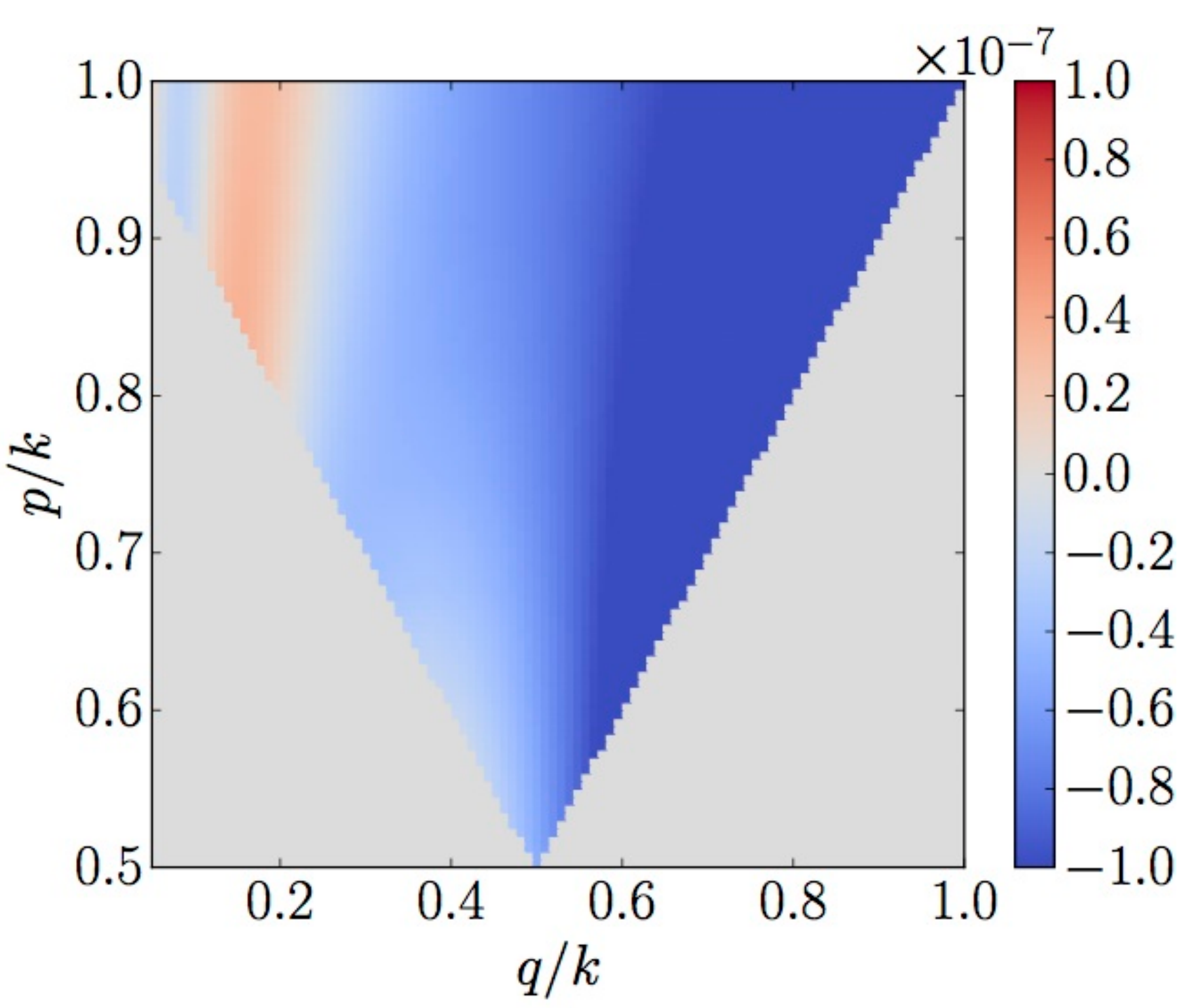} 
 \includegraphics[width=7.2cm]
{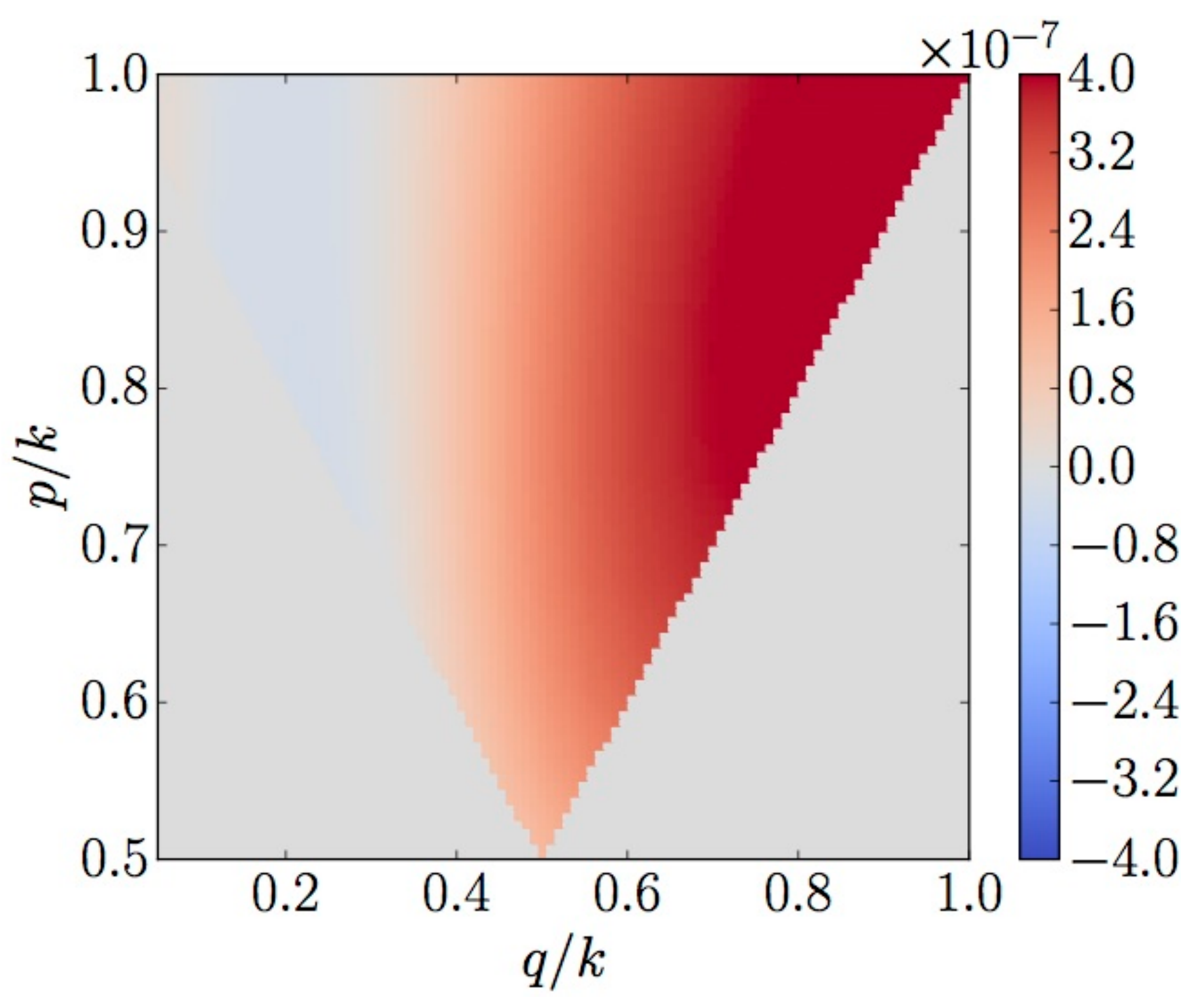} \hspace{0.4cm}
 \includegraphics[width=7.2cm]
{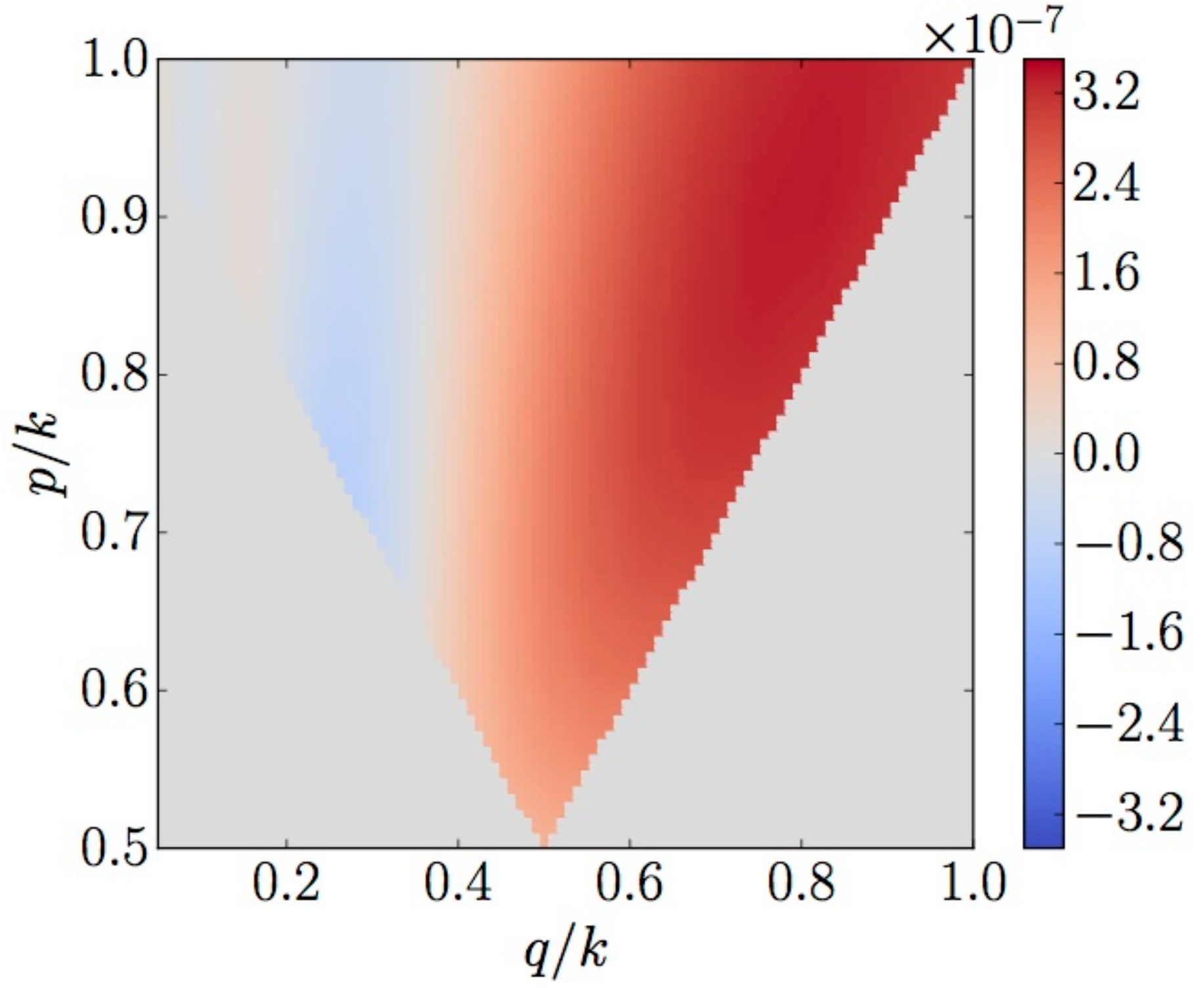} 
 \caption
 {Plot of the contributions to $(pqk)^2\mathcal{B}_{hh\zeta}$ arising from the $\mathcal{A}$, $\mathcal{B}$ and $\mathcal{C}$ terms in Eq.~(\ref{bisp_final}) (upper panels and lower left panel), and plot of the total $(pqk)^2\mathcal{B}_{hh\zeta}$ (lower right panel). The following set of parameters has been chosen in generating the plots above: $m_Q = 3.45$, $\epsilon_B = 3\times10^{-5}$, $\epsilon_{\phi} = 3\times10^{-3}$, $\epsilon_{\chi} = 3\times 10^{-8}$, $f = 10^{-2}\,M_{Pl}$,
$g = 10^{-2}$. We considered  these particular values for the  parameters for the sake of comparison with the existing literature. However, 
for a parameter region that has been filtered through the lenses of perturbativity bounds see \textit{Section}~\ref{results} and \textit{Appendix}~\ref{C}. Naturally, we expect that the shape, unlike the amplitude, does not depend on the choice of these parameters.}
 \label{fig2}
\end{center}
\end{figure}

\noindent We report in Fig.~\ref{fig2} the three contributions, (\ref{AA}) through (\ref{CC}), to the scalar-tensor-tensor bispectrum for a sample set of parameters. The sum of the three terms is also shown. As anticipated, the shape profile peaks in the equilateral configuration. \\Let us now move on to the bispectrum amplitude. It is instructive to report here the three contributions labelled $\mathcal{A},\mathcal{B},\mathcal{C}$:
\begin{eqnarray}\label{bisp_final}
\langle h^{(1)}\,h^{(1)}\,\zeta^{(2)\chi} \rangle_{\mathcal{A},\,\mathcal{B},\,\mathcal{C}}&\sim &\left(\frac{H}{M_{Pl}}\right)^3\,\left(\frac{U_{\chi}}{V_{\phi}}\right)\frac{\epsilon_{B}}{\sqrt{\epsilon_{\phi}}}\,\mathcal{I}_{\mathcal{A},\,\mathcal{B},\,\mathcal{C}}\\&\times&  \left(\int G^{\Psi}\cdot t\right)^2\left(\int G^{u}\cdot t\cdot t \right)\,,\nonumber
\end{eqnarray}
where
\begin{eqnarray}\label{Iabc}
\mathcal{I}_{\mathcal{A}}\equiv \frac{\lambda\,g\,\sqrt{\epsilon_{E}}\,M_{Pl}}{f}\,,\quad\quad\quad
\mathcal{I}_{\mathcal{B}}\equiv \frac{m_{Q}\,\lambda\,H}{f}\,,\quad\quad\quad
\mathcal{I}_{\mathcal{C}}\equiv \frac{\lambda\,H}{f}\,.
\end{eqnarray}

\begin{figure}[h!]
\begin{center}
  \includegraphics[width=7.2cm]{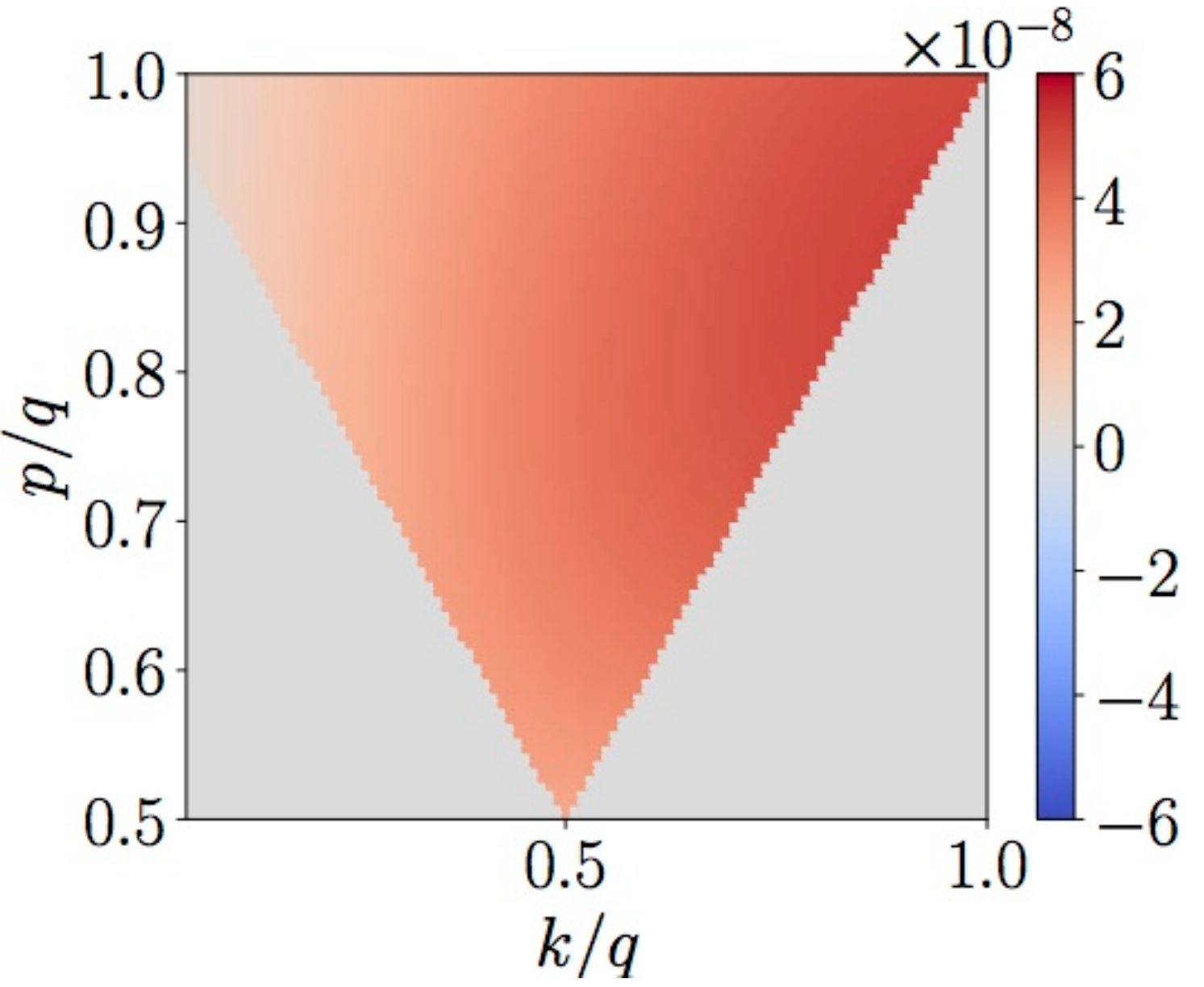}\hspace{0.4cm}
  \includegraphics[width=7.2cm]
{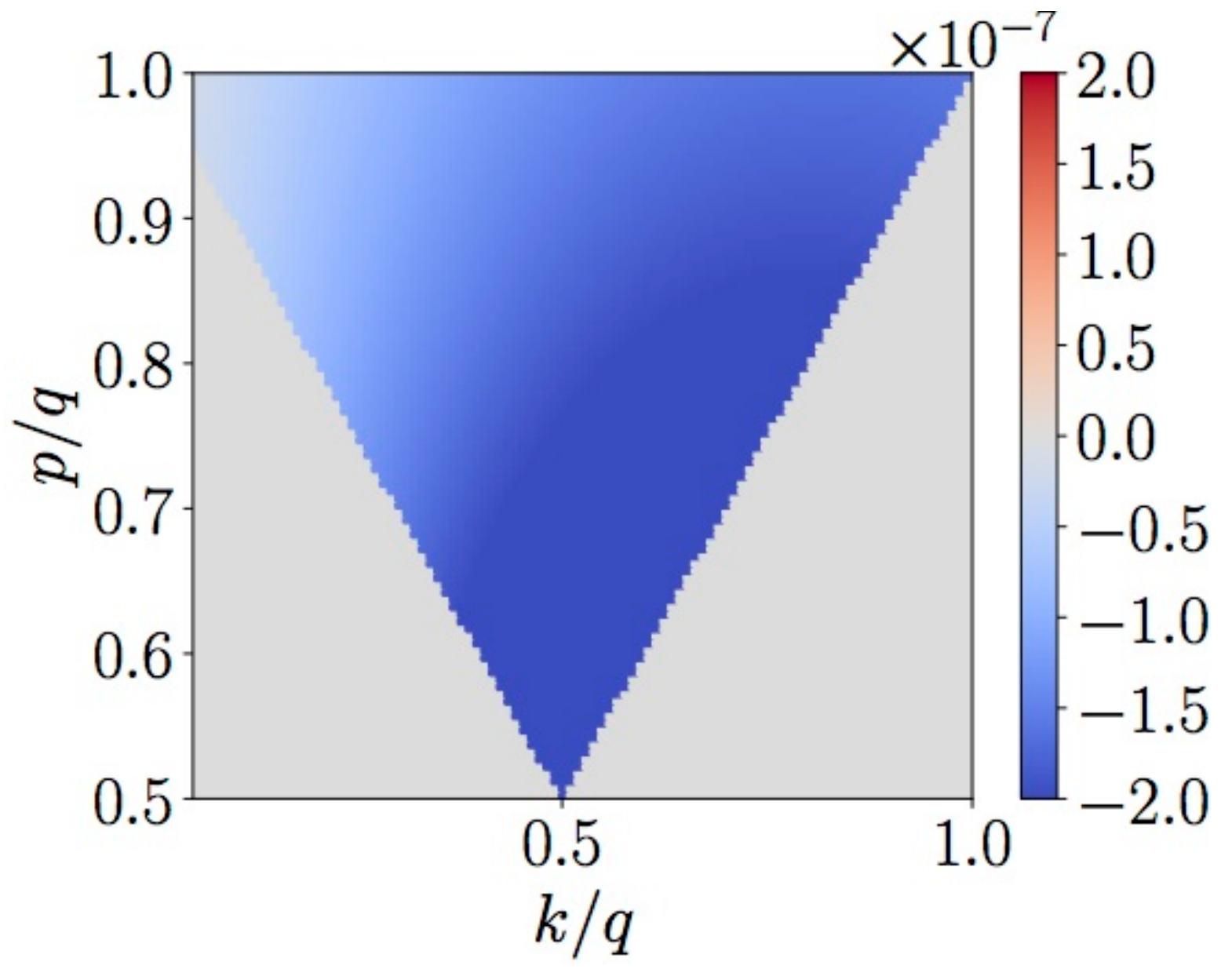} 
 \includegraphics[width=7.2cm]
{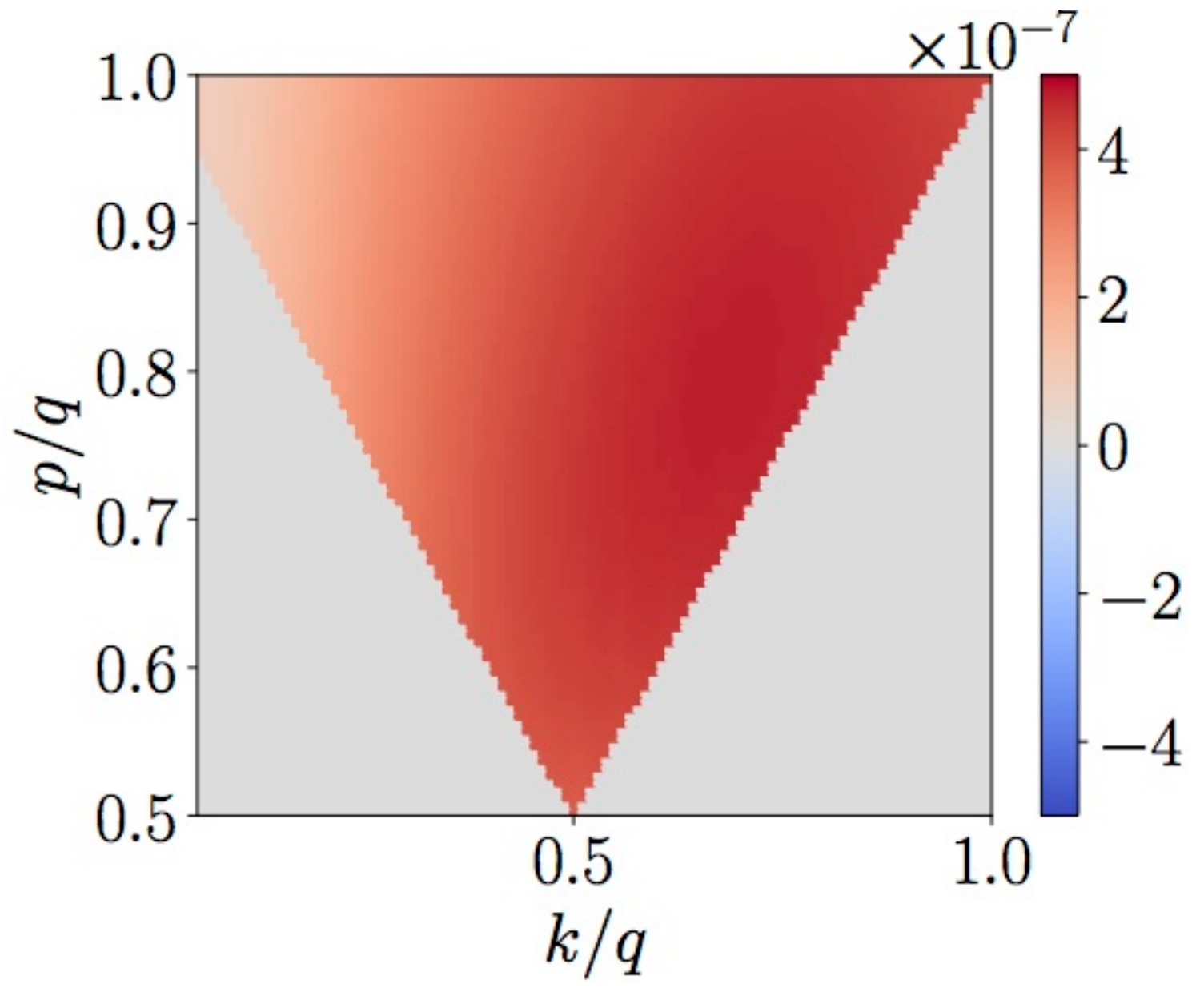} \hspace{0.4cm}
 \includegraphics[width=7.2cm]
{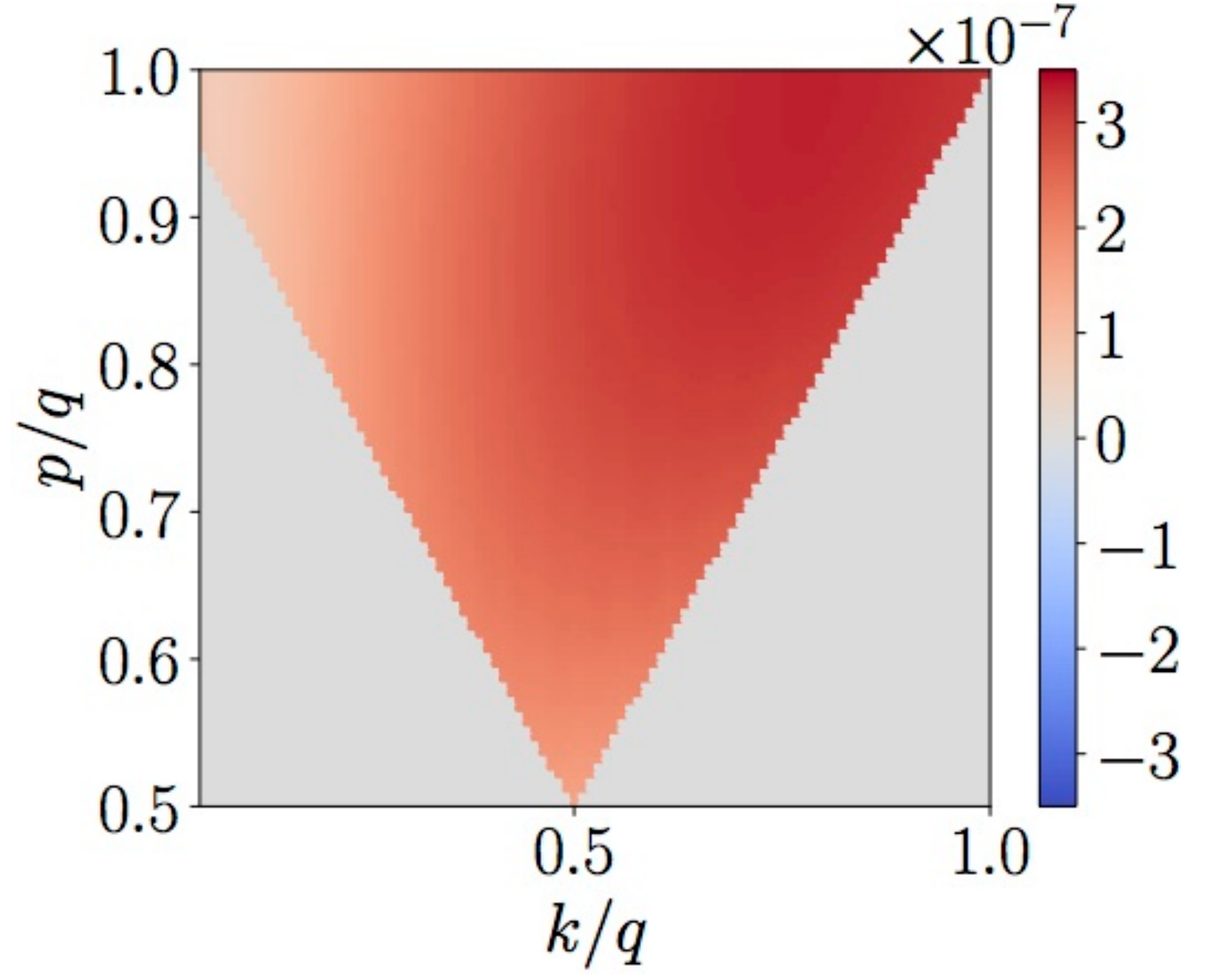} 
 \caption
 {Same as Fig.~\ref{fig2}, except for choosing as variables $k/q$ and $p/q$ (scalar fluctuation and one of the tensor fluctuation modes respectively).}
 \label{fig3}
\end{center}
\end{figure}
\noi Before elaborating further on the magnitude of the bispectrum, we take a quick detour to discuss the power spectrum contributions arising from the same interactions, i.e. $\langle \zeta^{(2)\chi}\,\zeta^{(2)\chi}\rangle$. This is a one-loop correction to the tree-level scalar power spectrum. The consistency of the perturbative expansion rests on the fact that such contribution, as well as those at higher loops, is sub-leading with respect to the tree-level observable. This fact will also be reflected on the bispectrum. Using Eqs.~(\ref{AA})-(\ref{CC}), one finds that the amplitude of the bispectrum (\ref{bisp_final}) can be expressed in the following form  
\begin{eqnarray}\label{ampl_one}
\langle h^{(1)}\,h^{(1)}\,\zeta^{(2)\chi} \rangle_{\mathcal{A},\,\mathcal{B},\,\mathcal{C}}&\sim & \mathcal{P}_{h}^{\text{s}} \left(\mathcal{P}_{\zeta}^{\text{tree}}\cdot\Delta^2\right)^{1/2}
\end{eqnarray}
where we have defined $\mathcal{P}_{\zeta}^{\text{tree}}$ as the scale-invariant tree-level power spectrum of curvature fluctuations and $\Delta^2$ as the correction introduced by the loop, i.e. $\langle \zeta^{(2)\chi}\,\zeta^{(2)\chi}\rangle\simeq \mathcal{P}_{\zeta}^{\text{tree}}\cdot\Delta^2$. In deriving (\ref{ampl_one}), the expression for $\mathcal{P}_{h}^{\text{s}}$ from Eq.~(\ref{psh}) has been used. Parametrizing the scalar-tensor-tensor non-Gaussianity as $f_{\text{nl}}\equiv \mathcal{B}_{hh\zeta}/\mathcal{P}_{\zeta}^2$, one obtains
\begin{eqnarray}\label{res_ampl}
f_{\text{nl}}^{\rm SU(2)}\sim r^2\cdot\frac{\mathcal{P}_{h}^{\text{s}}}{\mathcal{P}_{h}^{\text{tot}}}\left(\frac{\Delta^2}{r^2\,\mathcal{P}_{\zeta}^{\text{tree}}}\right)^{1/2}\,.
\end{eqnarray}
One may verify that, for $\Delta^2 <10^{-5/3}$, the parameter space of the model supports a one-loop contribution to the scalar power spectrum that is subdominant with respect to the tree-level contribution. Under the same condition, the scalar non-Gaussianity arising from these interactions  remains below the upper bounds from Planck and a sizable $\mathcal{P}_{h}^{\text{s}}\gtrsim \mathcal{P}_{h}^{\text{vacuum}}$ is allowed (see Appendix~\ref{C} for more details).

\noi It is useful at this stage to compare  the result in (\ref{res_ampl}) to the scalar-tensor-tensor non-Gaussianity in standard single field inflation. From \cite{Maldacena:2002vr}, and using the above definition for $f_{\text{nl}}$, one finds $f_{\text{nl}}^{\text{standard}}\sim r^2$. On the other hand, the parameter space of the model we have been studying allows for a bispectrum as large as
\begin{eqnarray}\label{res_ampl_s}
f_{\text{nl}}^{\text{gauge}}\sim \frac{10^{3}}{r} f_{\text{nl}}^{\text{standard}}\,.
\end{eqnarray}
The perturbativity bound notwithstanding, the  scalar-tensor-tensor non-Gaussianity from SU(2) gauge fields shows a remarkable enhancement with respect to the standard result.

%\noi {\rosso Any bounds so far or forecasts on this object? No}

\subsection{Tensor-scalar-scalar bispectrum}
\noi For the sake of completeness, we provide below an estimate of the SU(2) contribution to the tensor-scalar-scalar (tss) three-point function of the model. We leave a more thorough treatment to future work. There are two relevant diagrams contributing to the tss bispectrum that originate from the Chern-Simons interaction and from metric tensor-gauge field interactions:
\begin{figure}[ht]
\begin{center}
  \includegraphics[width=5.5cm]{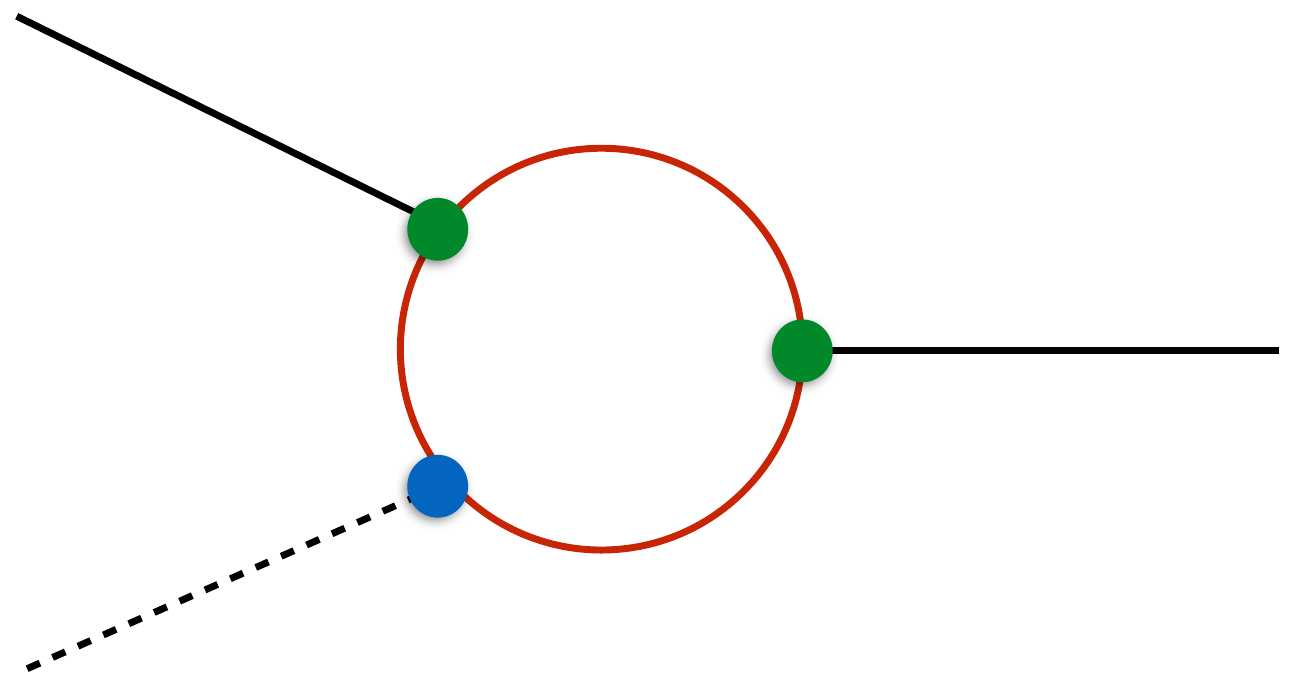}
  \hspace{2.0cm}
  \includegraphics[width=5.5cm]
{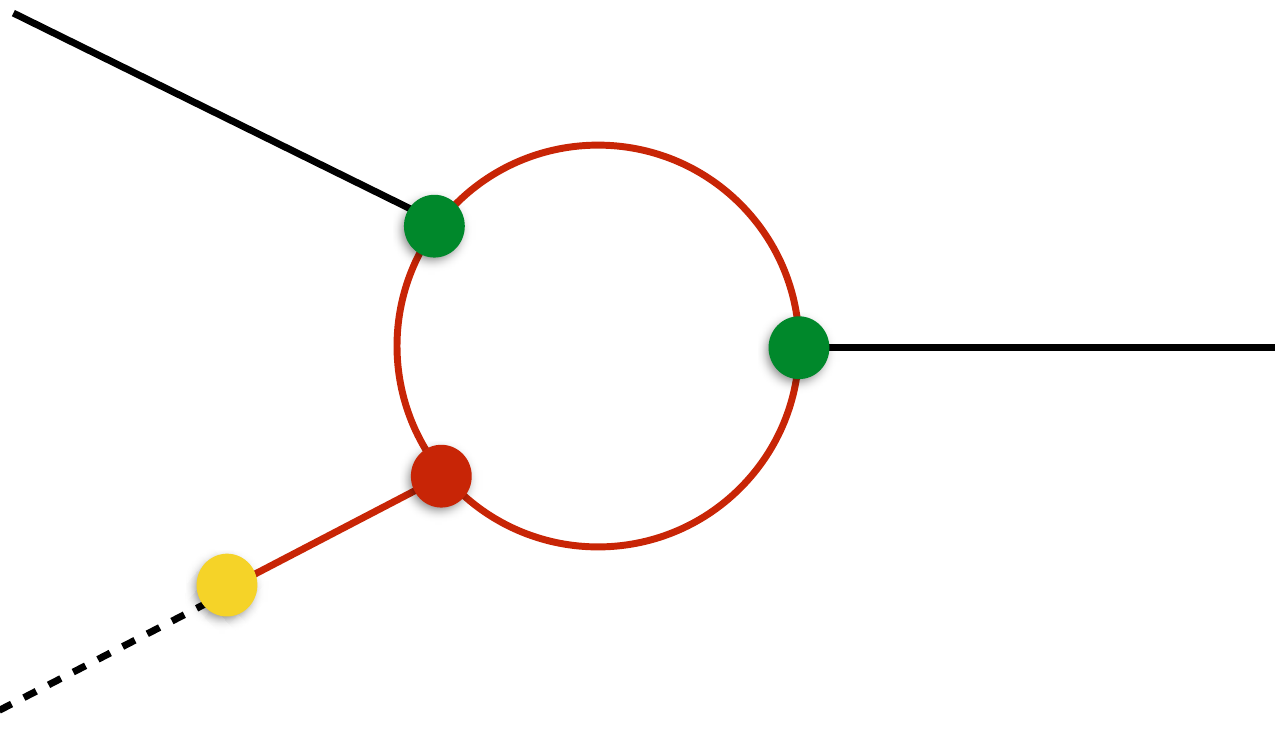} 
 \caption
 {Black dotted lines stand for propagators of the metric tensor perturbation $h\,$; red lines represent the gauge field tensor perturbation $t\,$; the solid black line indicates the curvature perturbation, $\zeta^{\chi}$. The green vertex stems from the Chern-Simons contribution to the $\delta\chi\, t^2 $ interaction, the yellow vertex from the quadratic Lagrangian in $h \, t$ (usual caveats apply), the blue and the red vertices from the cubic Lagrangian respectively in $h\,t^2$ and $t^3$.}
 \label{fig7}
\end{center}
\end{figure}

\noi Let us call (a) the diagram on the left and (b) the one of the right of Fig.~\ref{fig7}. We report below the estimate for the corresponding amplitudes:
\begin{eqnarray}\label{ampl}
\mathcal{B}_{h\zeta\zeta}^{(a)}&\approx& \mathcal{P}_{\zeta}^{1\text{loop}}\,c^{(ii)}\,\left(\frac{H}{M_{Pl}}\right)\,\left(t\cdot t\right)\simeq \mathcal{P}_{\zeta}^{1\text{loop}}\,m_{Q}\,\left(\frac{H}{M_{Pl}}\right)^2\,\left(t\cdot t\right)\simeq\mathcal{P}_{\zeta}^{1\text{loop}}\,\mathcal{P}_{h}^{\text{s}}\,\frac{m_{Q}}{\epsilon_{B}} \,,\nonumber\\
\\\label{ampl1}
\mathcal{B}_{h\zeta\zeta}^{(b)}&\approx& \mathcal{P}_{\zeta}^{1\text{loop}}\,\left(\mathcal{P}_{h}^{\text{s}}\right)^{1/2}\,c^{(i)}\,\left(t\cdot t\cdot t\right)\simeq \mathcal{P}_{\zeta}^{1\text{loop}}\,\left(\mathcal{P}_{h}^{\text{s}}\right)^{1/2}\,\frac{m_{Q}^{2}}{\sqrt{\epsilon_{B}}}\,\left(\frac{H}{M_{Pl}}\right)\,\left(t\cdot t\cdot t\right)\nonumber\\&\simeq& \mathcal{P}_{\zeta}^{1\text{loop}}\,\mathcal{P}_{h}^{\text{s}}\,m_{Q}\,\frac{m_{Q}}{\epsilon_{B}}\,\left(t\cdot t\right)\,,
\end{eqnarray}
where the coefficients $c^{(i)}$ and $c^{(ii)}$ are the coupling constants characterizing the cubic Lagrangian for tensor fluctuations (see Eqs.~(5)-(7) of \cite{Agrawal:2017awz}). In Eqs.~(\ref{ampl})-(\ref{ampl1}) we also used the fact that the SU(2)-sourced tensor power spectrum, $P_{h}^{\text{s}}$, is proportional to two (integrated) t mode-functions. \\

\noi Introducing $f_{\text{nl}}\equiv \mathcal{B}_{h\zeta\zeta}/\left(\mathcal{P}_{\zeta}\right)^2$, one finds
\begin{eqnarray}\label{fnla}
&& f_{\text{nl}}^{(a)}\sim r\,\left(\frac{\mathcal{P}_{h}^{\text{s}}}{\mathcal{P}_{h}^{\text{total}}}\right)\,\frac{\Delta^2\,m_{Q}}{\epsilon_{B}}  \,,\\\label{fnlb}
&& f_{\text{nl}}^{(b)}\sim  r\,\left(\frac{\mathcal{P}_{h}^{\text{s}}}{\mathcal{P}_{h}^{\text{total}}}\right)\,\frac{\Delta^2\,m_{Q}^2\,\left(t\cdot t\right)}{\epsilon_{B}} \,,
\end{eqnarray}
where $\mathcal{P}_{\zeta}^{1\text{loop}}\equiv \mathcal{P}_{\zeta}^{\text{tree}}\,\Delta^2$. The amplitude above are to be compared with the result from standard single field inflation \cite{Maldacena:2002vr}
\begin{eqnarray}
f_{\text{nl}}^{\text{standard}}\approx r  \,.
\end{eqnarray}
Taking the limiting values $\Delta^2=10^{-5/3}$, $\mathcal{P}_{h}^{\text{s}}=\mathcal{P}_{h}^{\text{total}}$ and setting $\left(t\cdot t\right)\sim e^{3.6\,m_{Q}}$, the amplitudes in Eqs.~(\ref{fnla})-(\ref{fnlb}) become
\begin{eqnarray}\label{fnla1}
&& f_{\text{nl}}^{(a)}\sim 10^{-2}\,r\cdot\frac{m_{Q}}{\epsilon_{B}}\approx 100\cdot r  \,,\\\label{fnlb1}
&& f_{\text{nl}}^{(b)}\sim  10^{-2}\,r\cdot\frac{m_{Q}^2\,e^{3.6\,m_{Q}}}{\epsilon_{B}} \approx 10^7\cdot r\,,
\end{eqnarray}
where in the last step the sample values $m_{Q}=3$, $\epsilon_{B}=10^{-4}$ have been used to provide a concrete comparison with the standard case. We pause here to stress that, unlike for the scalar-tensor-tensor bispectrum, the results in this subsection are to be considered  estimates and need to be confirmed by a full calculation. Since this observable is not the main focus of the paper, we leave a more thorough analysis to future work.

\section{Conclusions}
\label{conclusions}

The model studied here belongs to an important class of theories characterized by a sourced gravitational waves signal in excess of tensor vacuum fluctuations. The analysis of the dynamics and the signatures of similar set-ups represents a cautionary tale against the temptation to immediately read off the inflationary energy scale directly from the value of the tensor-to-scalar ratio.
%\\  
The distinctive signatures of the SU(2)-equipped model \cite{Dimastrogiovanni:2016fuu} includes a blue or otherwise bumpy chiral gravitational waves power spectrum to a level accessible by upcoming observations \cite{Dimastrogiovanni:2016fuu,Thorne:2017jft}, along with enhanced tensor non-Gaussianity \cite{Agrawal:2017awz}. Mixed tensor-scalar non-Gaussianities are just as important. These provide additional predictive power, which is crucial to help constrain the model parameters.
% \\

%\noi 
In this paper we derive predictions for the scalar-tensor-tensor bispectrum and focus in particular on the effects of the axion-SU(2) fields coupling. We find that the three-point function is significantly enhanced with respect to its counterpart in the minimal inflationary scenario. Our focus has been on the impact on observables of a controlled growth in the gauge tensor modes near horizon crossing. 
 Given that this dynamics is essentially localized  at the horizon, the resulting shape function is expected to peak in the equilateral configuration. This is indeed the outcome of our analysis, as shown in Figs.(\ref{fig2}-\ref{fig3}).
 % \\

%\noi 
The work presented here can be extended in a number of directions. It would be important to generate forecasts detailing the constraining power that upcoming experiments will have on mixed non-Gaussianity. A full analysis of the shape function also entails the comparison with existing templates  in order to help distinguish this class of models from other scenarios.

%\noi 
Our results call for detailed studies of the scalar sector of the theory resulting from the choice of a specific inflaton Lagrangian, $\mathcal{L}_{\phi}$. 
Perhaps most importantly, for a complete characterization of this and similar models it is essential to study the post-inflationary evolution of the axion and the gauge fields. 

%as a cross-check we plotted the bispectrum in the configuration in which the long mode is the one .... 

% contributions to tensor-tensor-scalar from fluctuations of the gauge field in the scalar sector (would probably be subdominant as the growth is experienced in the tensor sector)

\acknowledgments
ED and MF are delighted to thank E. Komatsu for illuminating conversations and kind encouragement. ED would like to thank the Perimeter Institute for Theoretical Physics (Canada) for hospitality and support whilst this work was in progress. ED is supported in part by DOE grant DE-SC0009946. HA, MF, KK and DW are supported by STFC grant ST/N000668/1. RJH is supported by UK Science and Technology Facilities Council grant ST/N5044245.  The work of KK has received funding from the European Research Council (ERC) under the European Union's Horizon 2020 research and innovation programme (grant agreement No.646702 ``CosTesGrav"). 

\begin{appendix}

\section{Derivation of the second-order equation of motion for $u$} 
\label{A}

\noi The equation of motion in real space, considering only the relevant source term, reads
\begin{equation}\label{}
u^{''}+\left(a^2 m_{\chi}^{2}-\partial^{2}-\frac{a^{''}}{a}\right) u=\frac{\lambda}{2 f}\left[\frac{d}{dt}\left(g\,a\,Q\,t^{2}_{ij}\right)+2\,\frac{d t_{ia}}{dt}\,\epsilon^{0ijk}\partial_{j} t_{ka}\right]\,,
\end{equation}
where $\partial^2\equiv \delta^{ij}\partial_{i}\partial_{j}$, $m_{\chi}^2\equiv d^2 U/d\chi^2 |_{\chi=\bar{\chi}}$ and $'\equiv d/d\tau$.\\

\noindent In momentum space one finds
\begin{eqnarray}\label{eq5}
u^{''}_{\textbf{k}}+\left(a^2 m_{\chi}^{2}+k^{2}-\frac{a^{''}}{a}\right) u_{\textbf{k}}&=&\frac{\lambda}{2 f}\int \frac{d^3 k_{2}}{(2\pi)^3}\sum_{\lambda_{1}\lambda_{2}}\Big\{e_{ij}^{\lambda_{1}}(\hat{k}_{1})e_{ij}^{\lambda_{2}}(\hat{k}_{2})\left[\frac{d}{dt}\left(g\,a\,Q\,t^{\lambda_{1}}_{\textbf{k}_{\textbf{1}}}\,t^{\lambda_{2}}_{\textbf{k}_{\textbf{2}}}\right)\right]\nonumber\\&+&e_{ia}^{\lambda_{1}}(\hat{k}_{1})e_{ka}^{\lambda_{2}}(\hat{k}_{2})\,2\,\epsilon^{0ijk}\,i\,k_{2j}\,\frac{d t_{\textbf{k}_{\textbf{1}}}^{\lambda_{1}}}{dt}\,t_{\textbf{k}_{\textbf{2}}}^{\lambda_{2}}\Big\}\,,
\end{eqnarray}
where $\textbf{k}_{\textbf{1}}\equiv \textbf{k}-\textbf{k}_{\textbf{2}}$. Using the relation $i\,\epsilon^{ijk}\,k_{i}\,e^{\lambda}_{j \ell}=\pm k e^{\lambda}_{k \ell}$, where $+$ is for $\lambda=L$ and $-$ is for $\lambda=R$, and the definitions for $\epsilon_{E}$ and $m_{Q}$, one arrives at Eq.~(\ref{eq6}).

% Eq.~(\ref{eq5}) becomes
%
%
%
%\begin{eqnarray}\label{eq5}
%u^{''}_{\textbf{k}}+\left(a^2 m_{\chi}^{2}+k^{2}-\frac{a^{''}}{a}\right) u_{\textbf{k}}&=&\frac{\lambda}{2 f}\int \frac{d^3 k_{2}}{(2\pi)^3}\sum_{\lambda_{1}\lambda_{2}}e_{ij}^{\lambda_{1}}(\hat{k}_{1})e_{ij}^{\lambda_{2}}(\hat{k}_{2})\Big\{\left[\textcolor{red}{-}\frac{d}{dt}\left(g\,a\,Q\,t^{\lambda_{1}}_{\textbf{k}_{\textbf{1}}}\,t^{\lambda_{2}}_{\textbf{k}_{\textbf{2}}}\right)\right]\nonumber\\&\textcolor{red}{\pm}&2\,k_{2}\,\frac{d t_{\textbf{k}_{\textbf{1}}}^{\lambda_{1}}}{dt}\,t_{\textbf{k}_{\textbf{2}}}^{\lambda_{2}}\Big\}\,.
%\end{eqnarray}
%
%
%

\section{$\langle{\Psi}^{(2)}{\Psi}^{(1)}{u}^{(1)}\rangle$ computation} 
\label{B}

\noindent We present here our derivation of the contribution from $\langle{\Psi}_{\textbf{q}}^{(2)}{\Psi}_{\textbf{p}}^{(1)}{u}_{\textbf{k}}^{(1)}\rangle$ to the scalar-tensor-tensor correlation:
\begin{eqnarray}
\langle {h}_{\textbf{q}}^{(2)}(\tau){h}_{\textbf{p}}^{(1)}(\tau){\zeta}_{\textbf{k}}^{(1)\chi}(\tau)\rangle=-\frac{4}{a^3(\tau)M_{Pl}^{2}}\left(\frac{U_{\chi}}{V_{\phi}}\right)\left(\frac{H}{\dot{\phi}}\right)\langle{\Psi}_{\textbf{q}}^{(2)}(\tau){\Psi}_{\textbf{p}}^{(1)}(\tau){u}_{\textbf{k}}^{(1)}(\tau)\rangle\,,
\end{eqnarray}
where
\begin{eqnarray}
{\Psi}_{\textbf{p}}^{(1)}(\tau)=\int_{-\infty}^{0}d\tau^{'}\mathcal{G}^{(\Psi)}_{p}(\tau,\tau^{'})\,D_{p}(\tau^{'}){t}_{p}^{(1)}(\tau^{'})\,,
\end{eqnarray}
and
\begin{eqnarray}
{u}_{\textbf{k}}^{(1)}(\tau)=a_{\textbf{k}}^{}u_{k}(\tau)+a_{\textbf{k}}^{\dagger}u_{k}^{*}(\tau)
\,,\quad\quad\quad u_{k}(\tau)=-\frac{1}{\sqrt{2\,k^3}\tau}(1+i k \tau)e^{-i k \tau}\,.
\end{eqnarray}
The metric fluctuation to second order reads
\begin{eqnarray}
{\Psi}_{\textbf{q}}^{(2)}(\tau)=\int d\tau^{'}\mathcal{G}^{\Psi}_{q}(\tau,\tau^{'})\,D_{q}(\tau^{'}){t}_{p}^{(2)}(\tau^{'})\,,
\end{eqnarray}
where 
\begin{eqnarray}
{t}^{(2)}_{\textbf{q}}(\tau^{'})=\int_{-\infty}^{\tau^{'}}d\tau^{''}\mathcal{G}_{q}^{t}(\tau^{'},\tau^{''}) \mathcal{J}^{(t)}_{\textbf{q}}(\tau^{''})\,,
\end{eqnarray}
and $\mathcal{G}_{q}^{t}$ is the Green's function for the gauge field tensor fluctuations. $\mathcal{J}^{(t)}_{\textbf{q}}$ is the source term appearing in the second-order equation of motion for the gauge field, specifically the one due to the $\delta\chi\cdot t\cdot t$ Chern-Simons interaction:
\begin{equation}
t^{''}_{\textbf{q}}+\left[k^2+\frac{2}{\tau^2}\left(1+m_{Q}^{2}+k\tau\left(2m_{Q}+m_{Q}^{-1}\right)\right)\right]t^{}_{\textbf{q}}=\mathcal{J}^{(t)}_{\textbf{q}}\,,
\end{equation}
where $'$ indicates the derivative w.r.t. conformal time $\tau$ and
\begin{eqnarray}
\mathcal{J}^{(t)}_{\textbf{q}}\equiv -\frac{\lambda}{f}\epsilon_{\alpha\beta}^{R}(-\hat{q})\int\frac{d^3 k_{1}}{(2\pi)^3}\sum_{\lambda=L,R}\epsilon_{\alpha\beta}^{\lambda}(\hat{k}_{1})\left[\left(\frac{gQ}{H\tau}\pm k_{1}\right)t^{\lambda}_{\textbf{k}_{1}}\,\delta\chi^{'}_{\textbf{k}_{2}}+\left(q\pm k_{1}\right)t^{\lambda\,'}_{\textbf{k}_{1}}\,\delta\chi^{}_{\textbf{k}_{2}}\right]\,.\nonumber\\
\end{eqnarray}
Here $\textbf{k}_{2}=\textbf{q}_{}-\textbf{k}_{1}$ and $\pm$ correspond, respectively, to $\lambda=L,\,R$. After performing the Wick contractions, one arrives at (for one permutation)
\begin{eqnarray}\label{inspec}
\langle\hat{\Psi}_{\textbf{q}}^{(2)}(\tau)\hat{\Psi}_{\textbf{p}}^{(1)}(\tau)\delta\hat{\chi}_{\textbf{k}}^{(1)}(\tau)\rangle&=&(2\pi)^3\delta^{(3)}(\textbf{q}+\textbf{p}+\textbf{k})\left(-\frac{\lambda}{f}\right)\nonumber\\&\times&\int_{-\infty}^{\tau}d\tau_{1}\,\mathcal{G}_{q}^{\Psi}(\tau,\tau_{1})D_{q}(\tau_{1})\int_{-\infty}^{\tau_{1}}d\tau_{2}\,\mathcal{G}_{q}^{t}(\tau_{1},\tau_{2})\nonumber\\&\times&a(\tau)\left[\frac{\left(p+q\right)^{2}-k^2}{4 p q}\right]^2\int_{-\infty}^{\tau}d\tau_{3}\,\mathcal{G}_{p}^{\Psi}(\tau,\tau_{3})D_{p}(\tau_{3})\nonumber\\&\times&\Big[\left(\frac{gQ}{H\tau_{2}}-p\right)\delta\chi^{'}_{k}(\tau_{2})\delta\chi^{*}_{k}(\tau)t^{R}_{p}(\tau_{2})t^{R*}_{p}(\tau_{3})\nonumber\\&+&\left(q-p\right)\delta\chi^{}_{k}(\tau_{2})\delta\chi^{*}_{k}(\tau)t^{R'}_{p}(\tau_{2})t^{R*}_{p}(\tau_{3}) \Big]\,.
\end{eqnarray}
%
%where we used
%
%\begin{equation}
%\epsilon^{L}_{\alpha\beta}(\hat{q})\epsilon^{L}_{\alpha\beta}(\hat{p})=\left[\frac{\left(p+q\right)^{2}-k^2}{4 p q}\right]^2\,,
%\end{equation}
%
% along with
 %
 %\begin{equation}
%i\,\epsilon^{ijk}k_{i}\epsilon^{(R,L)}_{jl}(\hat{k})=\mp k \epsilon^{(R,L)}_{kl}(\hat{k})\,,\quad\quad\quad \epsilon^{0ijk}=\epsilon^{ijk}\,.
%\end{equation}
%

\section{Bounds from perturbativity and from scalar non-Gaussianity} 
\label{C}

\noindent We estimate here the one-loop power spectrum arising from the same interactions contributing to the tensor-tensor-scalar bispectrum analyzed in this paper:
\begin{figure}[ht]
\begin{center}
  \includegraphics[width=5.6cm]{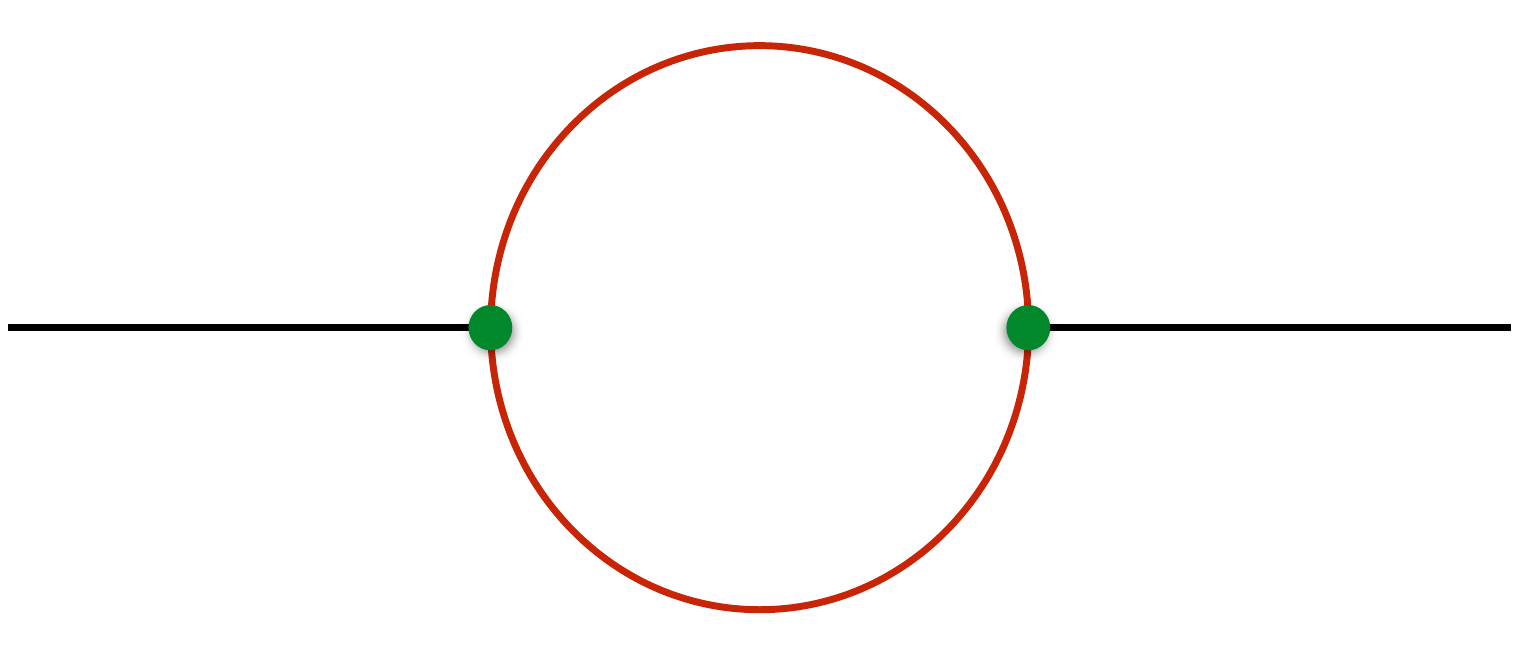}
\caption{Diagrammatic representation of $P_{\zeta}^{\text{1loop}}$ arising from the Chern-Simons interaction. Red lines represent the gauge field tensor perturbation, $t$, solid black lines represents the curvature perturbation, $\zeta$. The green vertex arises from the Chern-Simons contribution to $\delta\chi\cdot t \cdot t$.
}
 \label{fig4}
\end{center}
\end{figure}

\noi One finds (schematically)
\begin{eqnarray}\label{look}
\mathcal{P}_{\zeta}^{1\text{loop}}\simeq \left(\frac{H}{\sqrt{\epsilon_{\phi}}\,M_{Pl}}\right)^2\,\left(\frac{U_{\chi}}{V_{\phi}}\right)^2\left(\mathcal{I}_{\mathcal{A}}+
\mathcal{I}_{\mathcal{B}}+
\mathcal{I}_{\mathcal{C}}\right)^2\left(\int G^{u}\cdot t\cdot t \right)^2\,, 
\end{eqnarray}
where $\mathcal{I}_{\mathcal{A},\,\mathcal{B},\,\mathcal{C}}$ were introduced in Eqs.~(\ref{Iabc}) and we defined $\langle \zeta^{(2)\chi}_{\textbf{k}}\,\zeta^{(2)\chi}_{\textbf{q}}\rangle\simeq (2\pi)^3\delta^{(3)}(\textbf{k}_{}+\textbf{q})(k^3/2\pi^2)\mathcal{P}_{\zeta}^{1\text{loop}}(k)$. Let us now take a closer look at the parameters in the model to put (\ref{look}) in a more explicit form. To this aim, we will make use of Eqs.~(\ref{sol-sr1})-(\ref{sol-sr2}), which we report below
\begin{equation}\label{rewritt}
Q_{}=\left(\frac{-f\, U_{\chi}}{3g\lambda H}\right)^{1/3}\,,\quad\quad\frac{\lambda}{2fH}\dot{\chi}\simeq m_{Q}+\frac{1}{m_{Q}}\,.
\end{equation}
We also remind the reader that $m_{Q}\equiv g\,Q/H$. In the slow-roll regime for $Q$, one finds $\epsilon_{E}\approx\epsilon_{B}/m_{Q}^{2}$. From (\ref{rewritt}) and from the definition of the slow-roll parameters (see Sec.~\ref{model}) it follows that $(\lambda\,M_{Pl}/f)^2=(m_{Q}+m_{Q}^{-1})^2/\epsilon_{\chi}$. One also finds $g=(m_{Q}^{2}\,H)/(\sqrt{\epsilon_{B}}\,M_{Pl})$. From the field equations, assuming a standard background equation of motion for the inflaton, one also has
\begin{equation}
\frac{U_{\chi}}{V_{\phi}}\approx \frac{\lambda\,M_{Pl}}{f}\sqrt{\frac{\epsilon_{E}\epsilon_{B}}{\epsilon_{\phi}}}\,.
\end{equation}
Equipped with all of the above, one finds $\mathcal{P}_{\zeta}^{1\text{loop}}\approx\mathcal{P}_{\zeta}^{\text{tree}}\cdot\Delta^2$ where, schematically, we define $\Delta^2=(\Delta_{\mathcal{A}}+\Delta_{\mathcal{B}}+\Delta_{\mathcal{C}})^2$, with 
\begin{equation}
\Delta_{i}\approx 10^4\cdot e^{3.6\,m_{Q}}\cdot \frac{\epsilon_{B}}{\epsilon_{\chi}\epsilon_{\phi}}\left(m_{Q}+m_{Q}^{-1}\right)^{2}\left(\frac{H}{M_{Pl}}\right)^{2}\times \begin{cases}
    1,& i=\mathcal{A},\,\mathcal{}B\\
   m_{Q}^{-1},              & i=\mathcal{C}
\end{cases} 
\end{equation}
It is straightforward to verify that the bound from scalar non-Gaussianity, i.e. from $\langle \zeta^{(2)\chi}_{\textbf{k}_{}}\,\zeta^{(2)\chi}_{\textbf{q}}\,\zeta^{(2)\chi}_{\textbf{p}}\rangle\propto (\mathcal{P}_{\zeta}^{\text{1loop}})^{3/2}$, is given by $\Delta^2\lesssim10^{-5/3}$. In deriving the latter bound we considered current Planck constraints on equilateral non-Gaussianity. The saturation of this bound, which is slightly more stringent than the perturbativity bound on the power spectrum, corresponds to the amplitude reported in Eq.~(\ref{res_ampl_s}).

\end{appendix}


\begin{thebibliography}{95}

{\footnotesize
%\cite{Guth:1980zm}
\bibitem{inflation} 
  A.~H.~Guth,
  %``The Inflationary Universe: A Possible Solution to the Horizon and Flatness Problems,''
  %Phys.\ Rev.\ D {\bf 23}, 347 (1981).
 [\href{http://inspirehep.net/record/154280}{{\blu{ Phys.\ Rev.\ D {\bf 23}, 347 (1981)}}}];
 % doi:10.1103/PhysRevD.23.347
  %%CITATION = doi:10.1103/PhysRevD.23.347;%%
  %6824 citations counted in INSPIRE as of 08 Jun 2018
%\cite{Lyth:1998xn}
%\bibitem{Lyth:1998xn} 
  D.~H.~Lyth and A.~Riotto,
  %``Particle physics models of inflation and the cosmological density perturbation,''
  Phys.\ Rept.\  {\bf 314}, 1 (1999)
%  doi:10.1016/S0370-1573(98)00128-8
 % [hep-ph/9807278].
 [\href{https://arxiv.org/abs/hep-ph/9807278}{{\blu{hep-ph/9807278}}}];
  %%CITATION = doi:10.1016/S0370-1573(98)00128-8;%%
  %1624 citations counted in INSPIRE as of 08 Jun 2018
%\cite{Riotto:2002yw}
%\bibitem{Riotto:2002yw} 
  A.~Riotto,
  %``Inflation and the theory of cosmological perturbations,''
  ICTP Lect.\ Notes Ser.\  {\bf 14}, 317 (2003)
  [\href{https://arxiv.org/abs/hep-ph/0210162}{{\blu{hep-ph/0210162}}}];
  %[hep-ph/0210162].
  %%CITATION = HEP-PH/0210162;%%
  %272 citations counted in INSPIRE as of 08 Jun 2018
%\cite{Kinney:2003xf}
%\bibitem{Kinney:2003xf} 
  W.~H.~Kinney,
  %``Cosmology, inflation, and the physics of nothing,''
  NATO Sci.\ Ser.\ II {\bf 123}, 189 (2003)
 % doi:10.1007/978-94-010-0076-5_5
  %[astro-ph/0301448].
 [\href{https://arxiv.org/abs/astro-ph/0301448}{{\blu{astro-ph/0301448}}}];
  %%CITATION = doi:10.1007/978-94-010-0076-5_5;%%
  %62 citations counted in INSPIRE as of 08 Jun 2018
  %\cite{Wands:2007bd}
%\bibitem{Wands:2007bd} 
  D.~Wands,
  %``Multiple field inflation,''
  Lect.\ Notes Phys.\  {\bf 738}, 275 (2008)
%  doi:10.1007/978-3-540-74353-8_8
[\href{https://arxiv.org/abs/astro-ph/0702187}{{\blu arXiv:0702187}}];
%\cite{Baumann:2009ds}
%\bibitem{Baumann:2009ds} 
  D.~Baumann,
  %``Inflation,''
 % doi:10.1142/9789814327183_0010
  %arXiv:0907.5424 [hep-th].
  [\href{https://arxiv.org/abs/0907.5424}{{\blu{arXiv:0907.5424}}}].
  %%CITATION = doi:10.1142/9789814327183_0010;%%
  %498 citations counted in INSPIRE as of 08 Jun 2018






\bibitem{observations}
%\cite{Planck:2013jfk}
%\bibitem{Planck:2013jfk} 
  P.~A.~R.~Ade {\it et al.} [Planck Collaboration],
  %``Planck 2013 results. XXII. Constraints on inflation,''
  Astron.\ Astrophys.\  {\bf 571}, A22 (2014)
  %doi:10.1051/0004-6361/201321569
%  [arXiv:1303.5082 [astro-ph.CO]].
   [\href{https://arxiv.org/abs/1303.5082}{{\blu{arXiv:1303.5082}}}];
  %%CITATION = doi:10.1051/0004-6361/201321569;%%
  %1552 citations counted in INSPIRE as of 08 Jun 2018
%\cite{Ade:2013ydc}
%\bibitem{Ade:2013ydc} 
  P.~A.~R.~Ade {\it et al.} [Planck Collaboration],
  %``Planck 2013 Results. XXIV. Constraints on primordial non-Gaussianity,''
  Astron.\ Astrophys.\  {\bf 571}, A24 (2014)
  %doi:10.1051/0004-6361/201321554
 % [arXiv:1303.5084 [astro-ph.CO]].
   [\href{https://arxiv.org/abs/1303.5084}{{\blu{arXiv:1303.5084}}}];
  %%CITATION = doi:10.1051/0004-6361/201321554;%%
  %668 citations counted in INSPIRE as of 08 Jun 2018
  %\cite{Ade:2015ava}
%\bibitem{Ade:2015ava} 
  P.~A.~R.~Ade {\it et al.} [Planck Collaboration],
  %``Planck 2015 results. XVII. Constraints on primordial non-Gaussianity,''
  Astron.\ Astrophys.\  {\bf 594}, A17 (2016)
%  doi:10.1051/0004-6361/201525836
  %[arXiv:1502.01592 [astro-ph.CO]];
   [\href{http://arxiv.org/abs/1502.01592}{{\sf 1502.01592}}];
  %%CITATION = doi:10.1051/0004-6361/201525836;%%
  %482 citations counted in INSPIRE as of 12 Jun 2018
%\cite{Ade:2015lrj}
%\cite{Ade:2015lrj}
%\bibitem{Ade:2015lrj} 
  P.~A.~R.~Ade {\it et al.} [Planck Collaboration],
  %``Planck 2015 results. XX. Constraints on inflation,''
  Astron.\ Astrophys.\  {\bf 594}, A20 (2016)
 % doi:10.1051/0004-6361/201525898
     [\href{http://arxiv.org/abs/1502.02114}{{\sf 1502.02114}}].
%  [arXiv:1502.02114 [astro-ph.CO]].
  %%CITATION = doi:10.1051/0004-6361/201525898;%%
  %1655 citations counted in INSPIRE as of 13 Jun 2018

  


\bibitem{fossils}
%\cite{Giddings:2011zd}
%\bibitem{Giddings:2011zd} 
 % S.~B.~Giddings and M.~S.~Sloth,
  %``Cosmological observables, IR growth of fluctuations, and scale-dependent anisotropies,''
%  Phys.\ Rev.\ D {\bf 84}, 063528 (2011)
 % doi:10.1103/PhysRevD.84.063528
  %%CITATION = doi:10.1103/PhysRevD.84.063528;%%
  %60 citations counted in INSPIRE as of 06 Aug 2016
%\cite{Dai:2013ikl}
%\cite{Jeong:2012df}
%\bibitem{Jeong:2012df} 
  D.~Jeong and M.~Kamionkowski,
  %``Clustering Fossils from the Early Universe,''
  Phys.\ Rev.\ Lett.\  {\bf 108}, 251301 (2012)
 % doi:10.1103/PhysRevLett.108.251301
       [\href{http://arxiv.org/abs/1203.0302}{{\sf arXiv:1203.0302}}];
  %%CITATION = doi:10.1103/PhysRevLett.108.251301;%%
  %49 citations counted in INSPIRE as of 08 Jun 2018
  %\cite{Assassi:2012zq}
%\bibitem{Assassi:2012zq} 
  V.~Assassi, D.~Baumann and D.~Green,
  %``On Soft Limits of Inflationary Correlation Functions,''
  JCAP {\bf 1211}, 047 (2012)
%  doi:10.1088/1475-7516/2012/11/047
    [\href{https://arxiv.org/abs/1204.4207}{{\blu{arXiv:1204.4207}}}];
%\bibitem{Dai:2013ikl} 
  L.~Dai, D.~Jeong and M.~Kamionkowski,
  %``Seeking Inflation Fossils in the Cosmic Microwave Background,''
  Phys.\ Rev.\ D {\bf 87}, no. 10, 103006 (2013)
%  doi:10.1103/PhysRevD.87.103006
       [\href{http://arxiv.org/abs/1302.1868}{{\sf arXiv:1302.1868}}];
  %%CITATION = doi:10.1103/PhysRevD.87.103006;%%
  %16 citations counted in INSPIRE as of 06 Aug 2016
%\cite{Dai:2013kra}
%\bibitem{Dai:2013kra} 
  L.~Dai, D.~Jeong and M.~Kamionkowski,
  %``Anisotropic imprint of long-wavelength tensor perturbations on cosmic structure,''
  Phys.\ Rev.\ D {\bf 88}, no. 4, 043507 (2013)
%  doi:10.1103/PhysRevD.88.043507
         [\href{http://arxiv.org/abs/1306.3985}{{\sf arXiv:1306.3985}}];
%  [arXiv:1306.3985 [astro-ph.CO]].
  %%CITATION = doi:10.1103/PhysRevD.88.043507;%%
  %25 citations counted in INSPIRE as of 06 Aug 2016
%\cite{Brahma:2013rua}
%\bibitem{Brahma:2013rua} 
  S.~Brahma, E.~Nelson and S.~Shandera,
  %``Fossilized Gravitational Wave Relic and Primordial Clocks,''
  Phys.\ Rev.\ D {\bf 89}, no. 2, 023507 (2014)
%  doi:10.1103/PhysRevD.89.023507
           [\href{http://arxiv.org/abs/1310.0471}{{\sf arXiv:1310.0471}}];
%  [arXiv:1310.0471 [astro-ph.CO]].
  %%CITATION = doi:10.1103/PhysRevD.89.023507;%%
  %10 citations counted in INSPIRE as of 06 Aug 2016
%\cite{Dimastrogiovanni:2014ina}
%\bibitem{Dimastrogiovanni:2014ina} 
  E.~Dimastrogiovanni, M.~Fasiello, D.~Jeong and M.~Kamionkowski,
  %``Inflationary tensor fossils in large-scale structure,''
  JCAP {\bf 1412}, 050 (2014)
%  doi:10.1088/1475-7516/2014/12/050
             [\href{http://arxiv.org/abs/1407.8204}{{\sf arXiv:1407.8204}}];
%  [arXiv:1407.8204 [astro-ph.CO]].
  %%CITATION = doi:10.1088/1475-7516/2014/12/050;%%
  %15 citations counted in INSPIRE as of 06 Aug 2016
%\cite{Dimastrogiovanni:2015pla}
%\bibitem{Dimastrogiovanni:2015pla} 
  E.~Dimastrogiovanni, M.~Fasiello and M.~Kamionkowski,
  %``Imprints of Massive Primordial Fields on Large-Scale Structure,''
  JCAP {\bf 1602}, 017 (2016)
 % doi:10.1088/1475-7516/2016/02/017
               [\href{http://arxiv.org/abs/1504.05993}{{\sf arXiv:1504.05993}}];
%  [arXiv:1504.05993 [astro-ph.CO]].
  %%CITATION = doi:10.1088/1475-7516/2016/02/017;%%
  %15 citations counted in INSPIRE as of 06 Aug 2016
%\cite{Emami:2015uva}
%\bibitem{Emami:2015uva} 
  R.~Emami and H.~Firouzjahi,
  %``Clustering Fossil from Primordial Gravitational Waves in Anisotropic Inflation,''
  JCAP {\bf 1510}, no. 10, 043 (2015)
%  doi:10.1088/1475-7516/2015/10/043
    [\href{http://arxiv.org/abs/1506.00958}{{\sf arXiv:1506.00958}}];
%  [arXiv:1506.00958 [astro-ph.CO]].
  %%CITATION = doi:10.1088/1475-7516/2015/10/043;%%
  %7 citations counted in INSPIRE as of 06 Aug 2016
  %\cite{Chen:2009we}
%\bibitem{Chen:2009we} 
  X.~Chen and Y.~Wang,
  %``Large non-Gaussianities with Intermediate Shapes from Quasi-Single Field Inflation,''
  Phys.\ Rev.\ D {\bf 81}, 063511 (2010)
  %doi:10.1103/PhysRevD.81.063511
  %[arXiv:0909.0496 [astro-ph.CO]].
   [\href{http://arxiv.org/abs/0909.0496}{{\blu arXiv:0909.0496}}];
  %%CITATION = doi:10.1103/PhysRevD.81.063511;%%
  %157 citations counted in INSPIRE as of 18 Jun 2018
 %\cite{Chen:2009zp}
%\bibitem{Chen:2009zp} 
  X.~Chen and Y.~Wang,
  %``Quasi-Single Field Inflation and Non-Gaussianities,''
  JCAP {\bf 1004}, 027 (2010)
  %doi:10.1088/1475-7516/2010/04/027
  [arXiv:0911.3380 [hep-th]].
   [\href{http://arxiv.org/abs/0911.3380}{{\blu arXiv:0911.3380}}];
  %%CITATION = doi:10.1088/1475-7516/2010/04/027;%%
  %258 citations counted in INSPIRE as of 18 Jun 2018
  %\cite{Lee:2016vti}
%\bibitem{Lee:2016vti}
  H.~Lee, D.~Baumann and G.~L.~Pimentel,
 % ``Non-Gaussianity as a Particle Detector,''
  JHEP {\bf 1612} (2016) 040
%  doi:10.1007/JHEP12(2016)040
   [\href{http://arxiv.org/abs/1607.03735}{{\blu arXiv:1607.03735}}];
  %%CITATION = doi:10.1007/JHEP12(2016)040;%%
  %41 citations counted in INSPIRE as of 13 May 2018
 %\cite{Palma:2017lww}
%\bibitem{Palma:2017lww} 
  G.~A.~Palma and W.~Riquelme,
  %``Axion excursions of the landscape during inflation,''
  Phys.\ Rev.\ D {\bf 96}, no. 2, 023530 (2017)
%  doi:10.1103/PhysRevD.96.023530
  %[arXiv:1701.07918 [hep-th]].
   [\href{https://arxiv.org/abs/1701.07918}{{\blu{arXiv:1701.07918}}}];
  %%CITATION = doi:10.1103/PhysRevD.96.023530;%%
  %4 citations counted in INSPIRE as of 28 Jun 2018
  %\cite{Biagetti:2017viz}
%\bibitem{Biagetti:2017viz} 
  M.~Biagetti, E.~Dimastrogiovanni and M.~Fasiello,
  %``Possible Signatures of Inflationary Particle Content: Spin-2 Fields,''
  JCAP {\bf 1710}, no. 10, 038 (2017)
 % doi:10.1088/1475-7516/2017/10/038
    [\href{https://arxiv.org/abs/1708.01587}{{\blu{arXiv:1708.01587}}}];
  %%CITATION = doi:10.1088/1475-7516/2017/10/038;%%
  %2 citations counted in INSPIRE as of 15 Nov 2017
  %\cite{Chen:2018uul}
%\bibitem{Chen:2018uul}
  X.~Chen, G.~A.~Palma, W.~Riquelme, B.~Scheihing Hitschfeld and S.~Sypsas,
  %``Landscape tomography through primordial non-Gaussianity,''
 % arXiv:1804.07315 [hep-th];
    [\href{https://arxiv.org/abs/1804.07315}{{\blu{arXiv:1804.07315}}}];
  %%CITATION = ARXIV:1804.07315;%%
  %1 citations counted in INSPIRE as of 28 Jun 2018
    %\cite{Dimastrogiovanni:2018uqy}
%\bibitem{Dimastrogiovanni:2018uqy} 
  E.~Dimastrogiovanni, M.~Fasiello and G.~Tasinato,
  %``Probing the inflationary particle content: extra spin-2 field,''
   [\href{https://arxiv.org/abs/1806.00850}{{\blu{arXiv:1806.00850}}}].
  %%CITATION = ARXIV:1806.00850;%%

  
  
%%da quii  
  

%\cite{Arkani-Hamed:2015bza}
\bibitem{Arkani-Hamed:2015bza} 
  N.~Arkani-Hamed and J.~Maldacena,
  %``Cosmological Collider Physics,''
      [\href{http://arxiv.org/abs/1503.08043}{{\sf arXiv:1503.08043}}].
  %%CITATION = ARXIV:1503.08043;%%
  %126 citations counted in INSPIRE as of 08 Jun 2018



\bibitem{naturalinflation}
%\cite{Freese:1990rb}
%\bibitem{Freese:1990rb} 
  K.~Freese, J.~A.~Frieman and A.~V.~Olinto,
  %``Natural inflation with pseudo - Nambu-Goldstone bosons,''
 % Phys.\ Rev.\ Lett.\  {\bf 65}, 3233 (1990),
%  doi:10.1103/PhysRevLett.65.3233
   [\href{http://inspirehep.net/record/299254}{{\sf Phys.\ Rev.\ Lett.\  {\bf 65}, 3233 (1990)}}];
  %%CITATION = doi:10.1103/PhysRevLett.65.3233;%%
  %712 citations counted in INSPIRE as of 06 Aug 2016
%\cite{Freese:2004un}
%\bibitem{Freese:2004un} 
  K.~Freese and W.~H.~Kinney,
  %``On: Natural inflation,''
  Phys.\ Rev.\ D {\bf 70}, 083512 (2004)
%  doi:10.1103/PhysRevD.70.083512
   [\href{http://arxiv.org/abs/hep-ph/0404012}{{\sf hep-ph/0404012}}];
 % [hep-ph/0404012].
  %%CITATION = doi:10.1103/PhysRevD.70.083512;%%
  %44 citations counted in INSPIRE as of 06 Aug 2016
%\bibitem{fmp}     
  %\cite{Savage:2006tr}
%\bibitem{Savage:2006tr} 
%  C.~Savage, K.~Freese and W.~H.~Kinney,
  %``Natural Inflation: Status after WMAP 3-year data,''
%  Phys.\ Rev.\ D {\bf 74}, 123511 (2006)
 % doi:10.1103/PhysRevD.74.123511
%     [\href{http://arxiv.org/abs/hep-ph/0609144}{{\sf hep-ph/0609144}}];
%  [hep-ph/0609144].
  %%CITATION = doi:10.1103/PhysRevD.74.123511;%%
  %59 citations counted in INSPIRE as of 06 Aug 2016
%\cite{Freese:2014nla}
%\bibitem{Freese:2014nla} 
  K.~Freese and W.~H.~Kinney,
  %``Natural Inflation: Consistency with Cosmic Microwave Background Observations of Planck and BICEP2,''
  JCAP {\bf 1503}, 044 (2015)
 % doi:10.1088/1475-7516/2015/03/044
           [\href{http://arxiv.org/abs/1403.5277}{{\sf arXiv:1403.5277}}].
%  [arXiv:1403.5277 [astro-ph.CO]].
  %%CITATION = doi:10.1088/1475-7516/2015/03/044;%%
  %67 citations counted in INSPIRE as of 06 Aug 2016  

%\cite{Pajer:2013fsa}
\bibitem{Pajer:2013fsa} 
  E.~Pajer and M.~Peloso,
  %``A review of Axion Inflation in the era of Planck,''
  Class.\ Quant.\ Grav.\  {\bf 30}, 214002 (2013)
%  doi:10.1088/0264-9381/30/21/214002
  [\href{http://arxiv.org/abs/1305.3557}{{\sf arXiv:1305.3557}}].
  %%CITATION = doi:10.1088/0264-9381/30/21/214002;%%
  %73 citations counted in INSPIRE as of 08 Jun 2018



\bibitem{fmpt}      
%\cite{Kallosh:1995hi}
%\bibitem{Kallosh:1995hi} 
  R.~Kallosh, A.~D.~Linde, D.~A.~Linde and L.~Susskind,
  %``Gravity and global symmetries,''
  Phys.\ Rev.\ D {\bf 52}, 912 (1995)
 % doi:10.1103/PhysRevD.52.912
       [\href{http://arxiv.org/abs/hep-th/9502069}{{\sf hep-th/9502069}}];
%  [hep-th/9502069].
  %%CITATION = doi:10.1103/PhysRevD.52.912;%%
  %217 citations counted in INSPIRE as of 06 Aug 2016      
 %\cite{Banks:2003sx}
%\bibitem{Banks:2003sx} 
  T.~Banks, M.~Dine, P.~J.~Fox and E.~Gorbatov,
  %``On the possibility of large axion decay constants,''
  JCAP {\bf 0306}, 001 (2003)
%  doi:10.1088/1475-7516/2003/06/001
        [\href{http://arxiv.org/abs/hep-th/0303252}{{\sf hep-th/0303252}}].
  %[hep-th/0303252].
  %%CITATION = doi:10.1088/1475-7516/2003/06/001;%%
  %188 citations counted in INSPIRE as of 06 Aug 2016     
      
 %\cite{Kim:2004rp}
\bibitem{Kim:2004rp} 
  J.~E.~Kim, H.~P.~Nilles and M.~Peloso,
  %``Completing natural inflation,''
  JCAP {\bf 0501}, 005 (2005)
 % doi:10.1088/1475-7516/2005/01/005
   [\href{http://arxiv.org/abs/hep-ph/0409138}{{\sf hep-ph/0409138}}];
   %\cite{Dimopoulos:2005ac}
%\bibitem{Dimopoulos:2005ac} 
  S.~Dimopoulos, S.~Kachru, J.~McGreevy and J.~G.~Wacker,
  %``N-flation,''
  JCAP {\bf 0808}, 003 (2008)
     [\href{https://arxiv.org/abs/hep-th/0507205}{{\sf hep-th/0507205}}];
%  doi:10.1088/1475-7516/2008/08/003
  %[hep-th/0507205].
  %%CITATION = doi:10.1088/1475-7516/2008/08/003;%%
  %480 citations counted in INSPIRE as of 12 Jun 2018
%\cite{Kallosh:2007cc}
%\bibitem{Kallosh:2007cc} 
  R.~Kallosh, N.~Sivanandam and M.~Soroush,
  %``Axion Inflation and Gravity Waves in String Theory,''
  Phys.\ Rev.\ D {\bf 77}, 043501 (2008)
    [\href{http://arxiv.org/abs/0710.3429}{{\sf  arXiv:0710.3429}}].
%  doi:10.1103/PhysRevD.77.043501
  %[arXiv:0710.3429 [hep-th]].
  %%CITATION = doi:10.1103/PhysRevD.77.043501;%%
  %71 citations counted in INSPIRE as of 12 Jun 2018




\bibitem{vast}
%\bibitem{Anber:2009ua} 
  M.~M.~Anber and L.~Sorbo,
%  ``Naturally inflating on steep potentials through electromagnetic dissipation,''
  Phys.\ Rev.\ D {\bf 81}, 043534 (2010)
    [\href{http://arxiv.org/abs/0908.4089}{{\sf  arXiv:0908.4089}}];
  %[arXiv:0908.4089 [hep-th]].
  %%CITATION = ARXIV:0908.4089;%%
%\cite{Sorbo:2011rz}
%\bibitem{Barnaby:2010vf} 
  N.~Barnaby and M.~Peloso,
%  ``Large Nongaussianity in Axion Inflation,''
  Phys.\ Rev.\ Lett.\  {\bf 106}, 181301 (2011)
    [\href{http://arxiv.org/abs/1011.1500}{{\sf  arXiv:1011.1500}}];
%  [arXiv:1011.1500 [hep-ph]].
  %%CITATION = ARXIV:1011.1500;%%
  %64 citations counted in INSPIRE as of 04 Nov 2014
%\bibitem{Sorbo:2011rz} 
  L.~Sorbo,
%  ``Parity violation in the Cosmic Microwave Background from a pseudoscalar inflaton,''
  JCAP {\bf 1106}, 003 (2011)
    [\href{http://arxiv.org/abs/1101.1525}{{\sf  arXiv:1101.1525}}];
 % [arXiv:1101.1525 [astro-ph.CO]].
  %%CITATION = ARXIV:1101.1525;%%
  %34 citations counted in INSPIRE as of 12 Feb 2014
%
%\cite{Barnaby:2011vw}
%\bibitem{Barnaby:2011vw} 
  N.~Barnaby, R.~Namba and M.~Peloso,
%  ``Phenomenology of a Pseudo-Scalar Inflaton: Naturally Large Nongaussianity,''
  JCAP {\bf 1104}, 009 (2011)
    [\href{http://arxiv.org/abs/1102.4333}{{\sf  arXiv:1102.4333}}];
  %[arXiv:1102.4333 [astro-ph.CO]].
  %%CITATION = ARXIV:1102.4333;%%
  %49 citations counted in INSPIRE as of 04 Nov 2014
%\cite{Cook:2011hg}
%\bibitem{Cook:2011hg} 
  J.~L.~Cook and L.~Sorbo,
%  ``Particle production during inflation and gravitational waves detectable by ground-based interferometers,''
  Phys.\ Rev.\ D {\bf 85}, 023534 (2012)
  [Erratum-ibid.\ D {\bf 86}, 069901 (2012)]
    [\href{http://arxiv.org/abs/1109.0022}{{\sf  arXiv:1109.0022}}];
  %[arXiv:1109.0022 [astro-ph.CO]].
  %%CITATION = ARXIV:1109.0022;%%
  %39 citations counted in INSPIRE as of 27 Sep 2014
%\cite{Barnaby:2012xt}
%\bibitem{Barnaby:2012xt} 
  N.~Barnaby, J.~Moxon, R.~Namba, M.~Peloso, G.~Shiu and P.~Zhou,
%  ``Gravity waves and non-Gaussian features from particle production in a sector gravitationally coupled to the inflaton,''
  Phys.\ Rev.\ D {\bf 86}, 103508 (2012)
    [\href{http://arxiv.org/abs/1206.6117}{{\sf  arXiv:1206.6117}}];
 % [arXiv:1206.6117 [astro-ph.CO]].
  %%CITATION = ARXIV:1206.6117;%%
  %33 citations counted in INSPIRE as of 27 Sep 2014
%\cite{Mukohyama:2014gba}
%\bibitem{Mukohyama:2014gba} 
  S.~Mukohyama, R.~Namba, M.~Peloso and G.~Shiu,
 % ``Blue Tensor Spectrum from Particle Production during Inflation,''
  JCAP {\bf 1408}, 036 (2014)
    [\href{http://arxiv.org/abs/1405.0346}{{\sf  arXiv:1405.0346}}];
  %[arXiv:1405.0346 [astro-ph.CO]].
  %%CITATION = ARXIV:1405.0346;%%
  %10 citations counted in INSPIRE as of 28 Sep 2014
 %\cite{Ferreira:2014zia}
%\bibitem{Ferreira:2014zia} 
  R.~Z.~Ferreira and M.~S.~Sloth,
  %``Universal Constraints on Axions from Inflation,''
  JHEP {\bf 1412}, 139 (2014)
 % doi:10.1007/JHEP12(2014)139
   [\href{http://arxiv.org/abs/1409.5799}{{\sf  arXiv:1409.5799}}];
 % [arXiv:1409.5799 [hep-ph]].
  %%CITATION = doi:10.1007/JHEP12(2014)139;%%
  %32 citations counted in INSPIRE as of 06 Aug 2016 %\cite{Ozsoy:2014sba}
 %\cite{Ozsoy:2014sba}
%\bibitem{Ozsoy:2014sba} 
  O.~\"{O}zsoy, K.~Sinha and S.~Watson,
  %``How Well Can We Really Determine the Scale of Inflation?,''
  Phys.\ Rev.\ D {\bf 91}, no. 10, 103509 (2015)
 % doi:10.1103/PhysRevD.91.103509
  %[arXiv:1410.0016 [hep-th]];
   [\href{http://arxiv.org/abs/1410.0016}{{\sf  arXiv:1410.0016}}];
  %%CITATION = doi:10.1103/PhysRevD.91.103509;%%
  %24 citations counted in INSPIRE as of 12 Jun 2018
  %\cite{Bartolo:2014hwa}
%\bibitem{Bartolo:2014hwa} 
  N.~Bartolo, S.~Matarrese, M.~Peloso and M.~Shiraishi,
  %``Parity-violating and anisotropic correlations in pseudoscalar inflation,''
  JCAP {\bf 1501}, no. 01, 027 (2015)
     [\href{http://arxiv.org/abs/1411.2521}{{\sf  arXiv:1411.2521}}];
%  doi:10.1088/1475-7516/2015/01/027
 % [arXiv:1411.2521 [astro-ph.CO]]
  %%CITATION = doi:10.1088/1475-7516/2015/01/027;%%
  %28 citations counted in INSPIRE as of 12 Jun 2018
  %\cite{Adshead:2015pva}
%\bibitem{Adshead:2015pva} 
  P.~Adshead, J.~T.~Giblin, T.~R.~Scully and E.~I.~Sfakianakis,
  %``Gauge-preheating and the end of axion inflation,''
  JCAP {\bf 1512}, no. 12, 034 (2015)
%  doi:10.1088/1475-7516/2015/12/034
  %[arXiv:1502.06506 [astro-ph.CO]];
   [\href{http://arxiv.org/abs/1502.06506}{{\sf  arXiv:1502.06506}}];
  %%CITATION = doi:10.1088/1475-7516/2015/12/034;%%
  %43 citations counted in INSPIRE as of 12 Jun 2018
 %\cite{Bartolo:2015dga}
%\bibitem{Bartolo:2015dga} 
  N.~Bartolo, S.~Matarrese, M.~Peloso and M.~Shiraishi,
  %``Parity-violating CMB correlators with non-decaying statistical anisotropy,''
  JCAP {\bf 1507}, no. 07, 039 (2015)
   [\href{http://arxiv.org/abs/1505.02193}{{\sf  arXiv:1505.02193}}];
%  doi:10.1088/1475-7516/2015/07/039
  %[arXiv:1505.02193 [astro-ph.CO]].
  %%CITATION = doi:10.1088/1475-7516/2015/07/039;%%
  %26 citations counted in INSPIRE as of 12 Jun 2018
% \bibitem{Namba:2015gja} 
  R.~Namba, M.~Peloso, M.~Shiraishi, L.~Sorbo and C.~Unal,
  %``Scale-dependent gravitational waves from a rolling axion,''
  JCAP {\bf 1601}, no. 01, 041 (2016)
 % doi:10.1088/1475-7516/2016/01/041
   [\href{http://arxiv.org/abs/1509.07521}{{\sf  arXiv:1509.07521}}];
%\bibitem{Ferreira:2015omg} 
  R.~Z.~Ferreira, J.~Ganc, J.~Noreña and M.~S.~Sloth,
  %``On the validity of the perturbative description of axions during inflation,''
  JCAP {\bf 1604}, no. 04, 039 (2016)
 % doi:10.1088/1475-7516/2016/04/039
  [\href{http://arxiv.org/abs/1512.06116}{{\sf  arXiv:1512.06116}}];
%  [arXiv:1512.06116 [astro-ph.CO]].
  %%CITATION = doi:10.1088/1475-7516/2016/04/039;%%
  %6 citations counted in INSPIRE as of 06 Aug 2016
%\cite{Peloso:2016gqs}
%\bibitem{Peloso:2016gqs} 
  M.~Peloso, L.~Sorbo and C.~Unal,
  %``Rolling axions during inflation: perturbativity and signatures,''
 [\href{http://arxiv.org/abs/1606.00459}{{\sf  arXiv:1606.00459}}];
   %arXiv:1606.00459 [astro-ph.CO].
  %%CITATION = ARXIV:1606.00459;%%
  %2 citations counted in INSPIRE as of 06 Aug 2016
  %\cite{Adshead:2016iae}
%\bibitem{Adshead:2016iae} 
  P.~Adshead, J.~T.~Giblin, T.~R.~Scully and E.~I.~Sfakianakis,
  %``Magnetogenesis from axion inflation,''
  JCAP {\bf 1610}, 039 (2016)
 % doi:10.1088/1475-7516/2016/10/039
  %[arXiv:1606.08474 [astro-ph.CO]].
     [\href{http://arxiv.org/abs/1606.08474}{{\sf  arXiv:1606.08474}}];
  %%CITATION = doi:10.1088/1475-7516/2016/10/039;%%
  %32 citations counted in INSPIRE as of 12 Jun 2018
  %\cite{Garcia-Bellido:2016dkw}
%\bibitem{Garcia-Bellido:2016dkw} 
  J.~Garcia-Bellido, M.~Peloso and C.~Unal,
  %``Gravitational waves at interferometer scales and primordial black holes in axion inflation,''
  JCAP {\bf 1612}, no. 12, 031 (2016)
   [\href{http://arxiv.org/abs/1610.03763}{{\sf  arXiv:1610.03763}}];
%  doi:10.1088/1475-7516/2016/12/031
  %[arXiv:1610.03763 [astro-ph.CO]],
  %%CITATION = doi:10.1088/1475-7516/2016/12/031;%%
  %38 citations counted in INSPIRE as of 12 Jun 2018
  %\cite{Ozsoy:2017blg}
%\bibitem{Ozsoy:2017blg} 
  O.~\"{O}zsoy,
  %``On Synthetic Gravitational Waves from Multi-field Inflation,''
  JCAP {\bf 1804}, no. 04, 062 (2018)
 % doi:10.1088/1475-7516/2018/04/062
  %[arXiv:1712.01991 [astro-ph.CO]].
     [\href{http://arxiv.org/abs/1712.01991}{{\sf  arXiv:1712.01991}}];
  %%CITATION = doi:10.1088/1475-7516/2018/04/062;%%
%\cite{Fujita:2018zbr}
%\bibitem{Fujita:2018zbr} 
  T.~Fujita, I.~Obata, T.~Tanaka and S.~Yokoyama,
  %``Statistically Anisotropic Tensor Modes from Inflation,''
%  arXiv:1801.02778 [astro-ph.CO].
   [\href{http://arxiv.org/abs/1801.02778}{{\sf  arXiv:1801.02778}}].
  %%CITATION = ARXIV:1801.02778;%%
  %1 citations counted in INSPIRE as of 12 Jun 2018





%\cite{Ferreira:2017lnd}
\bibitem{Ferreira:2017lnd} 
  R.~Z.~Ferreira and A.~Notari,
  %``Thermalized Axion Inflation,''
  JCAP {\bf 1709}, no. 09, 007 (2017)
%  doi:10.1088/1475-7516/2017/09/007
    [\href{http://arxiv.org/abs/1706.00373}{{\sf arXiv:1706.00373}}]; 
  %%CITATION = doi:10.1088/1475-7516/2017/09/007;%%
  %7 citations counted in INSPIRE as of 08 Jun 2018
  %\cite{Ferreira:2017wlx}
%\bibitem{Ferreira:2017wlx} 
  R.~Z.~Ferreira and A.~Notari,
  %``Thermalized axion inflation: natural and monomial inflation with small $r$,''
  Phys.\ Rev.\ D {\bf 97}, no. 6, 063528 (2018)
%  doi:10.1103/PhysRevD.97.063528
  [\href{http://arxiv.org/abs/1711.07483}{{\sf arXiv:1711.07483}}]; 
  %%CITATION = doi:10.1103/PhysRevD.97.063528;%%
  %1 citations counted in INSPIRE as of 08 Jun 2018




\bibitem{cs}
%\cite{Biagetti:2013kwa}
%\bibitem{Biagetti:2013kwa} 
  M.~Biagetti, M.~Fasiello and A.~Riotto,
  %``Enhancing Inflationary Tensor Modes through Spectator Fields,''
  Phys.\ Rev.\ D {\bf 88}, 103518 (2013)
  %doi:10.1103/PhysRevD.88.103518
  [\href{http://arxiv.org/abs/1305.7241}{{\sf arXiv:1305.7241}}]; 
  %[arXiv:1305.7241 [astro-ph.CO]].
  %%CITATION = doi:10.1103/PhysRevD.88.103518;%%
  %38 citations counted in INSPIRE as of 12 Jun 2018
%\cite{Biagetti:2014asa}
%\bibitem{Biagetti:2014asa} 
  M.~Biagetti, E.~Dimastrogiovanni, M.~Fasiello and M.~Peloso,
  %``Gravitational Waves and Scalar Perturbations from Spectator Fields,''
  JCAP {\bf 1504}, 011 (2015)
  %doi:10.1088/1475-7516/2015/04/011
  [\href{http://arxiv.org/abs/1411.3029}{{\sf arXiv:1411.3029}}]; 
 % [arXiv:1411.3029 [astro-ph.CO]].
  %%CITATION = doi:10.1088/1475-7516/2015/04/011;%%
  %23 citations counted in INSPIRE as of 12 Jun 2018
%\cite{Fujita:2014oba}
%\bibitem{Fujita:2014oba} 
  T.~Fujita, J.~Yokoyama and S.~Yokoyama,
  %``Can a spectator scalar field enhance inflationary tensor mode?,''
  PTEP {\bf 2015}, 043E01 (2015)
  [\href{http://arxiv.org/abs/1411.3658}{{\sf arXiv:1411.3658}}].
%  doi:10.1093/ptep/ptv037
  %[arXiv:1411.3658 [astro-ph.CO]].
  %%CITATION = doi:10.1093/ptep/ptv037;%%
  %15 citations counted in INSPIRE as of 12 Jun 2018




\bibitem{mg}
%\cite{Dubovsky:2009xk}
%\bibitem{Dubovsky:2009xk} 
  S.~Dubovsky, R.~Flauger, A.~Starobinsky and I.~Tkachev,
  %``Signatures of a Graviton Mass in the Cosmic Microwave Background,''
  Phys.\ Rev.\ D {\bf 81}, 023523 (2010)
   [\href{http://arxiv.org/abs/0907.1658}{{\sf arXiv:0907.1658}}];
%  doi:10.1103/PhysRevD.81.023523
  %[arXiv:0907.1658 [astro-ph.CO]].
  %%CITATION = doi:10.1103/PhysRevD.81.023523;%%
  %38 citations counted in INSPIRE as of 12 Jun 2018
%\cite{Cusin:2014psa}
%\bibitem{Cusin:2014psa} 
  G.~Cusin, R.~Durrer, P.~Guarato and M.~Motta,
  %``Gravitational waves in bigravity cosmology,''
  JCAP {\bf 1505}, no. 05, 030 (2015)
   [\href{http://arxiv.org/abs/1412.5979}{{\sf arXiv:1412.5979}}];
  %%CITATION = doi:10.1088/1475-7516/2015/05/030;%%
  %42 citations counted in INSPIRE as of 12 Jun 2018
%\cite{Fasiello:2015csa}
%\bibitem{Fasiello:2015csa} 
  M.~Fasiello and R.~H.~Ribeiro,
  %``Mild bounds on bigravity from primordial gravitational waves,''
  JCAP {\bf 1507}, no. 07, 027 (2015)
     [\href{http://arxiv.org/abs/1505.00404}{{\sf arXiv:1505.00404}}].
  %%CITATION = doi:10.1088/1475-7516/2015/07/027;%%
  %19 citations counted in INSPIRE as of 12 Jun 2018








%\cite{Caldwell:2017chz}
\bibitem{Caldwell:2017chz} 
  R.~R.~Caldwell and C.~Devulder,
  %``Axion Gauge Field Inflation and Gravitational Leptogenesis: A Lower Bound on B Modes from the Matter-Antimatter Asymmetry of the Universe,''
  Phys.\ Rev.\ D {\bf 97}, no. 2, 023532 (2018)
 % doi:10.1103/PhysRevD.97.023532
        [\href{http://arxiv.org/abs/1706.03765}{{\sf arXiv:1706.03765}}]; 
  %%CITATION = doi:10.1103/PhysRevD.97.023532;%%
  %10 citations counted in INSPIRE as of 08 Jun 2018


\bibitem{recentCNI}
 %\cite{Adshead:2012kp}
%\bibitem{allCNI} 
  P.~Adshead and M.~Wyman,
  %``Chromo-Natural Inflation: Natural inflation on a steep potential with classical non-Abelian gauge fields,''
  Phys.\ Rev.\ Lett.\  {\bf 108}, 261302 (2012)
%  doi:10.1103/PhysRevLett.108.261302
    [\href{http://arxiv.org/abs/1202.2366}{{\sf arXiv:1202.2366}}]; 
    %\cite{Adshead:2012qe}
%\bibitem{Adshead:2012qe} 
  P.~Adshead and M.~Wyman,
  %``Gauge-flation trajectories in Chromo-Natural Inflation,''
  Phys.\ Rev.\ D {\bf 86}, 043530 (2012)
 % doi:10.1103/PhysRevD.86.043530
   [\href{http://arxiv.org/abs/1203.2264}{{\sf arXiv:1203.2264}}]; 
 % [arXiv:1203.2264 [hep-th]].
  %%CITATION = doi:10.1103/PhysRevD.86.043530;%%
  %28 citations counted in INSPIRE as of 10 Aug 2016
  %\cite{Martinec:2012bv}
%\bibitem{Martinec:2012bv} 
  E.~Martinec, P.~Adshead and M.~Wyman,
  %``Chern-Simons EM-flation,''
  JHEP {\bf 1302}, 027 (2013)
%  doi:10.1007/JHEP02(2013)027
  [\href{http://arxiv.org/abs/1206.2889}{{\sf arXiv:1206.2889}}]; 
%  [arXiv:1206.2889 [hep-th]].
  %%CITATION = doi:10.1007/JHEP02(2013)027;%%
  %23 citations counted in INSPIRE as of 10 Aug 2016
%\cite{Dimastrogiovanni:2012st}
%\bibitem{Dimastrogiovanni:2012st} 
  E.~Dimastrogiovanni, M.~Fasiello and A.~J.~Tolley,
  %``Low-Energy Effective Field Theory for Chromo-Natural Inflation,''
  JCAP {\bf 1302}, 046 (2013)
%  doi:10.1088/1475-7516/2013/02/046
    [\href{http://arxiv.org/abs/1211.1396}{{\sf arXiv:1211.1396}}].



%\cite{Lozanov:2018kpk}
\bibitem{Lozanov:2018kpk} 
  K.~D.~Lozanov, A.~Maleknejad and E.~Komatsu,
  %``Schwinger Effect by an $SU(2)$ Gauge Field during Inflation,''
     [\href{http://arxiv.org/abs/1805.09318}{{\sf arXiv:1805.09318}}].
%  arXiv:1805.09318 [hep-th].
  %%CITATION = ARXIV:1805.09318;%%
  %1 citations counted in INSPIRE as of 12 Jun 2018



%\cite{Maleknejad:2011jw}
\bibitem{Maleknejad:2013npa} 
  A.~Maleknejad and M.~M.~Sheikh-Jabbari,
  %``Gauge-flation: Inflation From Non-Abelian Gauge Fields,''
  Phys.\ Lett.\ B {\bf 723}, 224 (2013)
 % doi:10.1016/j.physletb.2013.05.001
 % [arXiv:1102.1513 [hep-ph]].
   [\href{http://arxiv.org/abs/1102.1513}{{\sf arXiv:1102.1513}}];
  %%CITATION = doi:10.1016/j.physletb.2013.05.001;%%
  %118 citations counted in INSPIRE as of 18 Jun 2018
  %\cite{Maleknejad:2011sq}
%\bibitem{Maleknejad:2011sq} 
  A.~Maleknejad and M.~M.~Sheikh-Jabbari,
  %``Non-Abelian Gauge Field Inflation,''
  Phys.\ Rev.\ D {\bf 84}, 043515 (2011)
 % doi:10.1103/PhysRevD.84.043515
  %[arXiv:1102.1932 [hep-ph]].
     [\href{http://arxiv.org/abs/1102.1932}{{\sf arXiv:1102.1932}}];
  %%CITATION = doi:10.1103/PhysRevD.84.043515;%%
  %95 citations counted in INSPIRE as of 18 Jun 2018
%\cite{Maleknejad:2013npa}
%\bibitem{Maleknejad:2013npa} 
  A.~Maleknejad and E.~Erfani,
  %``Chromo-Natural Model in Anisotropic Background,''
  JCAP {\bf 1403}, 016 (2014)
 % doi:10.1088/1475-7516/2014/03/016
   [\href{http://arxiv.org/abs/1311.3361}{{\sf arXiv:1311.3361}}];
%  [arXiv:1311.3361 [hep-th]].
  %%CITATION = doi:10.1088/1475-7516/2014/03/016;%%
  %8 citations counted in INSPIRE as of 06 Aug 2016
%\cite{Bhattacharjee:2014toa}
%\bibitem{Bhattacharjee:2014toa} 
  A.~Bhattacharjee, A.~Deshamukhya and S.~Panda,
  %``A note on low energy effective theory of chromo-natural inflation in the light of BICEP2 results,''
  Mod.\ Phys.\ Lett.\ A {\bf 30}, no. 11, 1550040 (2015)
 % doi:10.1142/S0217732315500406
     [\href{http://arxiv.org/abs/1406.5858}{{\sf arXiv:1406.5858}}];
%  [arXiv:1406.5858 [astro-ph.CO]].
  %%CITATION = doi:10.1142/S0217732315500406;%%
%\cite{Obata:2014loa}
%\bibitem{Obata:2014loa} 
  I.~Obata, T.~Miura and J.~Soda,
  %``Chromo-Natural Inflation in the Axiverse,''
  Phys.\ Rev.\ D {\bf 92}, no. 6, 063516 (2015)
%  doi:10.1103/PhysRevD.92.063516
       [\href{http://arxiv.org/abs/1412.7620}{{\sf arXiv:1412.7620}}];
%  [arXiv:1412.7620 [hep-ph]].
  %%CITATION = doi:10.1103/PhysRevD.92.063516;%%
  %8 citations counted in INSPIRE as of 06 Aug 2016
%\cite{Obata:2016tmo}
%\bibitem{Obata:2016tmo} 
  I.~Obata {\it et al.},
  %``Chiral primordial gravitational waves from dilaton induced delayed chromonatural inflation,''
  Phys.\ Rev.\ D {\bf 93}, no. 12, 123502 (2016)
 % doi:10.1103/PhysRevD.93.123502
         [\href{http://arxiv.org/abs/1602.06024}{{\sf arXiv:1602.06024}}].
 % [arXiv:1602.06024 [hep-th]].
  %%CITATION = doi:10.1103/PhysRevD.93.123502;%%
  %8 citations counted in INSPIRE as of 06 Aug 2016




%\cite{Bielefeld:2014nza}
\bibitem{Bielefeld:2014nza} 
%\cite{Deskins:2013lfx}
%\bibitem{Deskins:2013lfx} 
  J.~T.~Deskins, J.~T.~Giblin and R.~R.~Caldwell,
  %``Gauge Field Preheating at the End of Inflation,''
  Phys.\ Rev.\ D {\bf 88}, no. 6, 063530 (2013)
%  doi:10.1103/PhysRevD.88.063530
  %[arXiv:1305.7226 [astro-ph.CO]].
        [\href{http://arxiv.org/abs/1305.7226}{{\sf arXiv:1305.7226}}];
  %%CITATION = doi:10.1103/PhysRevD.88.063530;%%
  %20 citations counted in INSPIRE as of 18 Jun 2018
%\cite{Maleknejad:2014wsa}
%\bibitem{Maleknejad:2014wsa} 
  A.~Maleknejad,
  %``Chiral Gravity Waves and Leptogenesis in Inflationary Models with non-Abelian Gauge Fields,''
  Phys.\ Rev.\ D {\bf 90}, no. 2, 023542 (2014)
%  doi:10.1103/PhysRevD.90.023542
 % [arXiv:1401.7628 [hep-th]].
      [\href{http://arxiv.org/abs/1401.7628}{{\sf arXiv:1401.7628}}];
  %%CITATION = doi:10.1103/PhysRevD.90.023542;%%
  %23 citations counted in INSPIRE as of 18 Jun 2018
  J.~Bielefeld and R.~R.~Caldwell,
  %``Chiral Imprint of a Cosmic Gauge Field on Primordial Gravitational Waves,''
  Phys.\ Rev.\ D {\bf 91}, no. 12, 123501 (2015)
%  doi:10.1103/PhysRevD.91.123501
    [\href{http://arxiv.org/abs/1412.6104}{{\sf arXiv:1412.6104}}];
%  [arXiv:1412.6104 [astro-ph.CO]].
  %%CITATION = doi:10.1103/PhysRevD.91.123501;%%
  %8 citations counted in INSPIRE as of 14 Aug 2016
%\cite{Bielefeld:2015daa}
%\bibitem{Bielefeld:2015daa} 
  J.~Bielefeld and R.~R.~Caldwell,
  %``Cosmological consequences of classical flavor-space locked gauge field radiation,''
  Phys.\ Rev.\ D {\bf 91}, no. 12, 124004 (2015)
%  doi:10.1103/PhysRevD.91.124004
    [\href{http://arxiv.org/abs/1503.05222}{{\sf arXiv:1503.05222}}];
%  [arXiv:1503.05222 [gr-qc]].
  %%CITATION = doi:10.1103/PhysRevD.91.124004;%%
  %3 citations counted in INSPIRE as of 14 Aug 2016
  %\cite{Maleknejad:2016dci}
%\bibitem{Maleknejad:2016dci} 
  A.~Maleknejad,
  %``Gravitational leptogenesis in axion inflation with SU(2) gauge field,''
  JCAP {\bf 1612}, no. 12, 027 (2016)
  %doi:10.1088/1475-7516/2016/12/027
  %[arXiv:1604.06520 [hep-ph]].
   [\href{http://arxiv.org/abs/1604.06520}{{\sf arXiv:1604.06520}}];
  %%CITATION = doi:10.1088/1475-7516/2016/12/027;%%
  %13 citations counted in INSPIRE as of 18 Jun 2018
%\cite{Caldwell:2016sut}
%\bibitem{Caldwell:2016sut} 
  R.~R.~Caldwell, C.~Devulder and N.~A.~Maksimova,
  %``Gravitational Wave - Gauge Field Oscillations,''
      [\href{http://arxiv.org/abs/1604.08939}{{\sf arXiv:1604.08939}}];
  %arXiv:1604.08939 [gr-qc].
  %%CITATION = ARXIV:1604.08939;%%
  %1 citations counted in INSPIRE as of 14 Aug 2016
%\cite{Caldwell:2017chz}
%\bibitem{Caldwell:2017chz} 
  R.~R.~Caldwell and C.~Devulder,
  %``Axion Gauge Field Inflation and Gravitational Leptogenesis: A Lower Bound on B Modes from the Matter-Antimatter Asymmetry of the Universe,''
  Phys.\ Rev.\ D {\bf 97}, no. 2, 023532 (2018)
        [\href{http://arxiv.org/abs/1706.03765}{{\sf arXiv:1706.03765}}].
 % doi:10.1103/PhysRevD.97.023532
  %[arXiv:1706.03765 [astro-ph.CO]].
  %%CITATION = doi:10.1103/PhysRevD.97.023532;%%
  %10 citations counted in INSPIRE as of 12 Jun 2018

  
  
%\cite{Dimastrogiovanni:2012ew}
\bibitem{Dimastrogiovanni:2012ew} 
  E.~Dimastrogiovanni and M.~Peloso,
  %``Stability analysis of chromo-natural inflation and possible evasion of Lyth?s bound,''
  Phys.\ Rev.\ D {\bf 87}, no. 10, 103501 (2013)
 % doi:10.1103/PhysRevD.87.103501
%  [arXiv:1212.5184 [astro-ph.CO]].
      [\href{http://arxiv.org/abs/1212.5184}{{\sf arXiv:1212.5184}}];
%\cite{Adshead:2013nka}
%\bibitem{Adshead:2013nka} 
  P.~Adshead, E.~Martinec and M.~Wyman,
  %``Perturbations in Chromo-Natural Inflation,''
  JHEP {\bf 1309}, 087 (2013)
%  doi:10.1007/JHEP09(2013)087
      [\href{http://arxiv.org/abs/1305.2930}{{\sf arXiv:1305.2930}}];
      %\cite{Namba:2013kia}
%\bibitem{Namba:2013kia} 
  R.~Namba, E.~Dimastrogiovanni and M.~Peloso,
  %``Gauge-flation confronted with Planck,''
  JCAP {\bf 1311}, 045 (2013)
 % doi:10.1088/1475-7516/2013/11/045
   [\href{http://arxiv.org/abs/1308.1366}{{\sf arXiv:1308.1366}}];
%  [arXiv:1308.1366 [astro-ph.CO]].
  %%CITATION = doi:10.1088/1475-7516/2013/11/045;%%
  %15 citations counted in INSPIRE as of 06 Aug 2016
      %\cite{Adshead:2013qp}
%\cite{Adshead:2013qp}
%\bibitem{Adshead:2013qp} 
  P.~Adshead, E.~Martinec and M.~Wyman,
  %``Gauge fields and inflation: Chiral gravitational waves, fluctuations, and the Lyth bound,''
  Phys.\ Rev.\ D {\bf 88}, no. 2, 021302 (2013)
%  doi:10.1103/PhysRevD.88.021302
     [\href{http://arxiv.org/abs/1301.2598}{{\sf arXiv:1301.2598}}].
 % [arXiv:1301.2598 [hep-th]].
  %%CITATION = doi:10.1103/PhysRevD.88.021302;%%
  %27 citations counted in INSPIRE as of 10 Aug 2016
    
  



%\cite{Dimastrogiovanni:2016fuu}
\bibitem{Dimastrogiovanni:2016fuu} 
  E.~Dimastrogiovanni, M.~Fasiello and T.~Fujita,
 % ``Primordial Gravitational Waves from Axion-Gauge Fields Dynamics,''
  JCAP {\bf 1701}, no. 01, 019 (2017)
 % doi:10.1088/1475-7516/2017/01/019
   [\href{https://arxiv.org/abs/1608.04216}{{\blu{arXiv:1608.04216}}}].
 % [arXiv:1608.04216 [astro-ph.CO]].
  %%CITATION = doi:10.1088/1475-7516/2017/01/019;%%
  %17 citations counted in INSPIRE as of 11 Jan 2018
  
  %\cite{Obata:2016tmo}
\bibitem{Obata:2016tmo} 
  I.~Obata {\it et al.},
  %``Chiral primordial Chiral primordial gravitational waves from dilaton induced delayed chromonatural inflation,''
  Phys.\ Rev.\ D {\bf 93}, no. 12, 123502 (2016)
 % Addendum: [Phys.\ Rev.\ D {\bf 95}, no. 10, 109903 (2017)]
%  doi:10.1103/PhysRevD.95.109903, 10.1103/PhysRevD.93.123502
     [\href{https://arxiv.org/abs/1602.06024}{{\blu{arXiv:1602.06024}}}].
  %%CITATION = doi:10.1103/PhysRevD.95.109903, 10.1103/PhysRevD.93.123502;%%
  %25 citations counted in INSPIRE as of 08 Jun 2018
  
  %\cite{DallAgata:2018ybl}
\bibitem{DallAgata:2018ybl} 
  G.~Dall'Agata,
  %``Chromo-Natural inflation in Supergravity,''
   [\href{http://arxiv.org/abs/1804.03104}{{\sf  arXiv:1804.03104}}];
 %%CITATION = ARXIV:1804.03104;%%
%\cite{McDonough:2018xzh}
%\bibitem{McDonough:2018xzh} 
  E.~McDonough and S.~Alexander,
  %``Observable Chiral Gravitational Waves from Inflation in String Theory,''
 % arXiv:1806.05684 [hep-th].
    [\href{http://arxiv.org/abs/1806.05684}{{\sf  arXiv:1806.05684}}].
  %%CITATION = ARXIV:1806.05684;%%



  %\cite{Thorne:2017jft}
\bibitem{Thorne:2017jft} 
  B.~Thorne, T.~Fujita, M.~Hazumi, N.~Katayama, E.~Komatsu and M.~Shiraishi,
  %``Finding the chiral gravitational wave background of an axion-SU(2) inflationary model using CMB observations and laser interferometers,''
  Phys.\ Rev.\ D {\bf 97}, no. 4, 043506 (2018)
%  doi:10.1103/PhysRevD.97.043506
           [\href{http://arxiv.org/abs/1707.03240}{{\sf  arXiv:1707.03240}}].
  %%CITATION = doi:10.1103/PhysRevD.97.043506;%%
  %7 citations counted in INSPIRE as of 04 May 2018

%\cite{Agrawal:2017awz}
\bibitem{Agrawal:2017awz} 
  A.~Agrawal, T.~Fujita and E.~Komatsu,
%  ``Large Tensor Non-Gaussianity from Axion-Gauge Fields Dynamics,''
 % arXiv:1707.03023 [astro-ph.CO].
   [\href{https://arxiv.org/abs/1707.03023}{{\blu{arXiv:1707.03023}}}];
  %%CITATION = ARXIV:1707.03023;%%
  %6 citations counted in INSPIRE as of 11 Jan 2018
  %\cite{Agrawal:2018mrg}
%\bibitem{Agrawal:2018mrg} 
  A.~Agrawal, T.~Fujita and E.~Komatsu,
  %``Tensor Non-Gaussianity from Axion-Gauge-Fields Dynamics : Parameter Search,''
     [\href{http://arxiv.org/abs/1802.09284}{{\sf  arXiv:1802.09284}}];
  %%CITATION = ARXIV:1802.09284;%%
  %1 citations counted in INSPIRE as of 04 May 2018
  %\cite{Agrawal:2018gzp}
%\bibitem{Agrawal:2018gzp} 
  A.~Agrawal,
  %``Non-Gaussianity of Inflationary Gravitational Waves from the Field Equation,''
 % arXiv:1804.01481 [astro-ph.CO].
     [\href{https://arxiv.org/abs/1804.01481}{{\blu{arXiv:1804.01481}}}].
  %%CITATION = ARXIV:1804.01481;%%
  %1 citations counted in INSPIRE as of 12 Jun 2018

  
  %\cite{Fujita:2017jwq}
\bibitem{Fujita:2017jwq} 
  T.~Fujita, R.~Namba and Y.~Tada,
  %``Does the detection of primordial gravitational waves exclude low energy inflation?,''
  Phys.\ Lett.\ B {\bf 778}, 17 (2018)
 % doi:10.1016/j.physletb.2017.12.014
       [\href{http://arxiv.org/abs/1705.01533}{{\sf  arXiv:1705.01533}}].
  %%CITATION = doi:10.1016/j.physletb.2017.12.014;%%
  %9 citations counted in INSPIRE as of 08 Jun 2018
  
  
  
  %\cite{Bartolo:2016ami}
\bibitem{lisa}
  N.~Bartolo {\it et al.},
 % ``Science with the space-based interferometer LISA. IV: Probing inflation with gravitational waves,''
  JCAP {\bf 1612} (2016) no.12,  026
%  doi:10.1088/1475-7516/2016/12/026
        [\href{http://arxiv.org/abs/1610.06481}{{\blu arXiv:1610.06481}}];
  %%CITATION = doi:10.1088/1475-7516/2016/12/026;%%
  %47 citations counted in INSPIRE as of 13 May 2018  
%\cite{Caprini:2018mtu}
%\bibitem{Caprini:2018mtu} 
  C.~Caprini and D.~G.~Figueroa,
  %``Cosmological Backgrounds of Gravitational Waves,''
  %arXiv:1801.04268 [astro-ph.CO].
          [\href{http://arxiv.org/abs/1801.04268}{{\blu arXiv:1801.04268}}].
  %%CITATION = ARXIV:1801.04268;%%
  %10 citations counted in INSPIRE as of 12 Jun 2018
    
  
  %\cite{Maldacena:2002vr}
\bibitem{Maldacena:2002vr} 
  J.~M.~Maldacena,
 % ``Non-Gaussian features of primordial fluctuations in single field inflationary models,''
  JHEP {\bf 0305}, 013 (2003)
%  doi:10.1088/1126-6708/2003/05/013
[\href{https://arxiv.org/abs/astro-ph/0210603}{{\blu{astro-ph/0210603}}}].
  %[astro-ph/0210603].
  %%CITATION = doi:10.1088/1126-6708/2003/05/013;%%
  %1752 citations counted in INSPIRE as of 16 Apr 2018





}




\end{thebibliography}
\end{document}